\documentstyle[graphicx,amstex]{elsart}

% ...

\newcommand{\eqnref}[1]{(\ref{#1})}
\newcommand{\llssor}{{\it ll}\/-SSOR}

\newcommand{\beq}{\begin{equation}}
\newcommand{\eeq}{\end{equation}}        
\newcommand{\bqa}{\begin{eqnarray}}        
\newcommand{\eqa}{\end{eqnarray}}        
\newcommand{\be}{\begin{enumerate}}
\newcommand{\ee}{\end{enumerate}}        
\newcommand{\bi}{\begin{itemize}}
\newcommand{\ei}{\end{itemize}}        

\newcommand{\eg}{{\frenchspacing e.\hspace{0.4mm}g.{}}}
\newcommand{\ie}{{\frenchspacing i.\hspace{0.4mm}e.{}}}
\newcommand{\eq}[1]{{\frenchspacing eq.~(\ref{#1})}}
\newcommand{\fig}[1]{{\frenchspacing Fig.~\ref{#1}}}
\newcommand{\nn}{\nonumber}

\newcommand{\one}{{\mathbf 1}}
\newcommand{\uno}{\one}
\newcommand{\Z}{Z \!\!\! Z}

\newcommand{\MathZ}{{\Bbb{Z}}}

\begin{document}

\begin{frontmatter}

\title{{\tiny
\hspace*{\fill} HLRZ-98/22\\
\hspace*{\fill} BUGHW-SC-98/4\\
\hspace*{\fill} WUP-98/20\\}
Preconditioning of Improved and ``Perfect'' Fermion Actions}

\author[HLRZ]{W.~Bietenholz}
\author[HLRZ]{N.~Eicker}
\author[WUPM]{A.~Frommer}
\author[WUPW]{Th.~Lippert}
\author[WUPM]{B.~Medeke}
\author[HLRZ]{K.~Schilling}
\author[WUPM]{G.~Weuffen}

\address[HLRZ]{HLRZ, c/o Research Center J\"{u}lich, D-52425
  J\"{u}lich, Germany}
\address[WUPM]{Department of Mathematics, University of
    Wuppertal, D-42097 Wuppertal, Germany}
\address[WUPW]{Department of Physics, University of
    Wuppertal, D-42097 Wuppertal, Germany}

\begin{abstract}
  We construct a {\em locally-lexicographic SSOR preconditioner} to
  accelerate the parallel iterative solution of linear systems of
  equations for two improved discretizations of lattice fermions:
  ({\em i\/}) the Sheikholeslami-Wohlert scheme where a non-constant
  block-diagonal term is added to the Wilson fermion matrix and {\em
    (ii)} renormalization group improved actions which incorporate
  couplings beyond nearest neighbors of the lattice fermion fields.
  In case ({\em i\/}) we find the block \llssor-scheme to be more
  effective by a factor $\approx 2$ than odd-even preconditioned
  solvers in terms of convergence rates, at $\beta=6.0$. For type
  ({\em ii\/}) actions, we show that our preconditioner accelerates
  the iterative solution of a linear system of hypercube fermions by a
  factor of 3 to 4.
\end{abstract}


\end{frontmatter}

\section{Introduction}

Traditionally, simulations of lattice quantum chromodynamics (QCD)
were based on near\-est-neighbor finite difference approximations of
the derivatives of classical fields.  It is a general observation made
in both quenched and full QCD that results from lattices with a
resolution $>0.1$ fm suffer from considerable discretization errors,
see \eg\ Ref.~\cite{GUP95}.  Even optimistic estimates expect an
increase in costs of full QCD simulations $\propto a^{-6}$ as the
lattice spacing $a$ is decreased \cite{LEP95}.\footnote{In extreme
  cases, in particular in QCD thermodynamics, the costs rise $\propto
  a^{-10}$ \cite{NIE96}.}.


The Wilson fermion formulation is appropriate with respect to flavor
symmetry, but plagued by discretization errors of ${\cal O}(a)$.
These effects have been found to be sizeable \eg\ in a compilation of
quenched world data for quark masses \cite{GUP96} and in the
determination of the renormalized quark mass, exploiting the PCAC
relation in the Schr\"odinger functional \cite{JAN95,JAN98}.

In lattice QCD, the extraction of physical continuum results requires
to approach two limits: the {\em chiral limit} defined as the point in
parameter space where the pion mass vanishes, and the {\em continuum
  limit} defined by vanishing lattice spacing $a$. The chiral limit
amounts to an increase of the inverse pion mass (the correlation
length $\xi_{\pi}$) in lattice units towards infinity. The lattice
volume, \ie\ the number of sites, must be increased accordingly, in
order to control finite size effects.  At this point, simulations of
QCD with dynamical fermions encounter the problem of solving the
fermionic linear system $MX=\phi$, where $M$ is the fermion matrix---a
compute intensive task.

The second step, moving towards the continuum limit, requires to
decrease the lattice spacing $a$.  The two issues are related: if one
would be able to get reliable results at larger lattice spacing, one
could avoid dealing with {\it prohibitively fine} physical lattice
resolutions on large physical volumes.  On the classical level, one
might just use higher order derivative terms in the fermion action in
order to push finite-$a$-effects to higher orders.  But quantum
effects will largely spoil the intended gains.

At present, there are two major trends to improve the fermionic
discretization.  One approach follows {\em Symanzik's on-shell
  improvement} program \cite{SYM83}. Irrelevant (dimension 5) counter
terms are added to both, lattice action and composite operators in
order to avoid ${\cal O}(a)$ artifacts. A particularly simple and
hence preferred scheme is the Sheikholeslami and Wohlert action (SWA)
\cite{SHE85}, where the Wilson action is modified by adding a local
term, the so-called clover term.  Hereby, the amount of storage is
doubled.  The clover term is sufficient, in principle, to cancel the
${\cal O}(a)$ errors in the action, provided that a constant $c_{SW}$
is tuned suitably.  The hope is to reach the continuum limit for a
given scaling quantity $Q$ as $Q(a)= Q_{continuum} + {\cal O}(a^2)$,
\ie\ without ${\cal O}(a)$ contamination. The development of a
non-perturbative tuning procedure has been documented in a series of
papers, see Refs.~\cite{JAN95,JAN98,LUE962,Lue97}.

Another promising ansatz is based on Wilson's renormalization group
\cite{WiKo}. It goes under the name {\em perfect actions}.  Perfect
lattice actions are located on renormalized trajectories in parameter
space that intersect the critical surface (at infinite correlation
length) in a fixed point of a renormalization group transformation.
By definition, perfect actions are free of any cut-off effects,
therefore they represent continuum physics at any lattice spacing $a$.

In practice, perfect actions can best be approximated for
asymptotically free theories starting from fixed point actions. Such
fixed point actions can be identified in a multi-grid procedure solely
by minimization, without performing functional integrals. Thus, the
task reduces to a classical field theory problem. The fixed point
action then serves as an approximation to a perfect action at finite
correlation length; this is a so-called classically perfect action
\cite{HAS94}.  However, even in this approximation the couplings
usually extend over an infinite range, so for practical purposes a
truncation to short distances is unavoidable.  In such schemes of
`truncated perfect actions' (TPA) one is forced to give up part of the
original quest for perfectness, for reasons of practicability.

It goes without saying that SWA and TPA can prove their full utility
only after combination with state-of-the-art solvers in actual
parallel implementations.  In the recent years, the inversion of the
standard Wilson fermion matrix could be accelerated considerably by
use of the BiCGStab algorithm \cite{FRO94} and novel parallel {\it
  l}\/ocally-{\it l}\/exicographic symmetric successive over-relaxation
(\llssor) preconditioning techniques \cite{FIS96}.

We start from the hypercube fermion (HF) approximation formulated for
free fermions in Ref.~\cite{BIE96}.  For an alternative variant, see
Ref.~\cite{DEG98}.  In order to meet the topological structure of TPA,
we shall follow a bottom-up approach by adding interactions to the
free fermion case through {\em hyper-links} within the unit-hypercube.
This results in 40 independent hyper-links per site and amounts to a
storage requirement five times as large as in SWA.

In general, the fermion matrix for both SWA and TPA can be written in
the generic form
\begin{equation}
M=D+A+B+\cdots,
\label{eq:shape}
\end{equation}
where $D$ represents diagonal blocks (containing $12\times 12$
sub-blocks), $A$ is a nearest-neighbor hopping term, while $B$ contains
next-to-nearest-neighbor couplings, and so on.  

The key point is that one can include into the \llssor\ process ({\em
  i}\/) the internal degrees of freedom of the block diagonal term $D$
as arising in SWA and ({\em ii}\/) 2-space, 3-space, and 4-space
hyper-links, as present in a TPA like HF.

After reviewing the fermionic matrices for SWA and HF in section 2, we
introduce {\em locally lexicographic} over-relaxation (\llssor)
preconditioning of SWA in section \ref{SSOR}.  We discuss three
variants for the diagonal blocks to be used in block SSOR.  In section
\ref{PARA}, we shall parallelize block SSOR within an extended
\llssor-scheme and we shall discuss the inclusion of the HF into this
framework. In section \ref{RESULTS}, we benchmark the block-\llssor\ 
preconditioner on SWA---for several values of $c_{SW}$---in
comparison with odd-even preconditioning.  Our testbed is a set of
quenched configurations on lattices of size $16^4$ at $\beta=6.0$ and
our implementation machines are a 32-node APE100/Quadrics and a SUN
Ultra workstation.  Using the HF on a quenched $8^4$ system, we
compare the SSOR preconditioned version with an non-preconditioned
one, for a variety of mass parameters, again at $\beta=6.0$.

\section{Improved Fermionic Actions}

In this section we briefly review the basics of SWA and TPA.  To fix
our notation, we write the fermionic lattice action as
\begin{equation}
S_{F} = \sum_{x,y} \bar \Psi_{x} M(x,y) \Psi_{y},
\end{equation}
where $M$ is the fermion matrix.

\subsection{Sheikholeslami-Wohlert Action}

For the Wilson fermion action (with Wilson parameter $r=1$),
supplemented by the Sheikholeslami-Wohlert term, we have
\begin{eqnarray}
  M_{SW}(x,y) &= & \Big[ \one + \frac{c_{SW}}{2}\kappa\sum_{\mu,\nu}
  \sigma_{\mu\nu} F_{\mu\nu}(x)
  \Big] \delta_{x,y}  \\
  &-& \kappa \sum_\mu \Big[ (\one-\gamma_\mu)U_\mu(x) \delta_{x,y-\hat
    \mu} + (\one+\gamma_\mu)U_{-\mu}(x) \delta_{x,y+\hat \mu}\Big],
  \nonumber
  \label{eq:clover_matrix}
\end{eqnarray}
where $\kappa$ is the standard Wilson hopping parameter, $c_{SW}$ is a
parameter that can be tuned to optimize ${\cal O}(a)$ cancellations,
and $\hat\mu$ is a unit vector.

The `local' clover term consists of $12\times 12$ diagonal blocks. Its
explicit structure in Dirac space is given by
\begin{equation}
  \frac{1}{2} \sum_{\mu,\nu}\sigma_{\mu\nu} F_{\mu\nu}
= 
  \left(
    \begin{array}{cccc}
      F_1&F_2&F_3&F_4\\
      F_2^\dagger&-F_1&F_4^\dagger&-F_3\\
      F_3&F_4&F_1&F_2\\
      F_4^\dagger&-F_3&F_2^\dagger&-F_1\\
    \end{array}
  \right) \ ,
  \label{eq:structure}
\end{equation}
with the entries $F_i$ being complex $3\times 3$ matrices,
which are shorthands for linear combinations of the $F_{\mu\nu}$,
\begin{equation}
\begin{array}{cc}
F_1=-F_{12}, &\quad F_2=-F_{23}-iF_{13},\\
F_3= F_{34}, &\quad F_4= F_{14}-iF_{24}.\\
\end{array}
\end{equation}
$F_{\mu\nu}$ is defined by 
\begin{equation}
F_{\mu\nu} :=\frac{i}{8}\left(
f_{\mu\nu}-
f^{\dagger}_{\mu\nu}
\right),
\end{equation}
where the clover term $f_{\mu\nu}$ reads
\begin{eqnarray*}
f_{\mu\nu}(x)  
& :=  & U_{\mu}(x)  U_{\nu}(x+\hat{\mu}) 
              U^{\dagger}_{\mu}(x+\hat{\nu})   U^{\dagger}_{\nu}(x)\\
& + & U_{\nu}(x)  U^{\dagger}_{\mu}(x-\hat{\mu}+\hat{\nu})  
          U^{\dagger}_{\nu}(x-\hat{\mu})  U_{\mu}(x-\hat{\mu})\\
& + & U^{\dagger}_{\mu}(x-\hat{\mu})  U^{\dagger}_{\nu}(x-\hat{\mu}-\hat{\nu})  
  U_{\mu}(x-\hat{\mu}-\hat{\nu})   U_{\nu}(x-\hat{\nu})\\
& + & U^{\dagger}_{\nu}(x-\hat{\nu})  U_{\mu}(x-\hat{\nu})  
  U_{\nu}(x+\hat{\mu}-\hat{\nu})   U^{\dagger}_{\mu}(x).\\
\end{eqnarray*}

The Wilson-Sheikholeslami-Wohlert matrix exhibits the well known
$\gamma_5$ symmetry
\begin{equation}
\gamma_5\,M_{SW}\gamma_5=M_{SW}^{\dagger},
\end{equation}
with the eigenvalues of $M_{SW}$ coming in complex-conjugate
pairs.

\subsection{Hypercube Fermions}

The physical properties of a given lattice action remain unaltered
under a block variable {\em renormalization group transformation}
(RGT). As a simple example, we can divide the (infinite) lattice into
disjoint hypercubic blocks of $n^{d}$ sites each and introduce new
variables living on the centers of these blocks (block factor $n$
RGT).  Then the RGT relates
\begin{equation}
\phi '_{x'} \sim \sum_{x\in x'} \phi_{x} ,
\end{equation}
where $\phi$ and $\phi '$ represent the original and the new lattice
fields, respectively. The points $x \in \MathZ^{d}$ are the sites of
the original lattice and $x'$ are those of the new lattice with
spacing $n$. $x\in x'$ means that the point $x$ belongs to the block
with center $x'$.

Now the original action $S[\phi ]$ transforms into a new action
$S'[\phi ']$ on the coarse lattice. The latter is determined by
the functional integral
\begin{equation}
e^{-S'[\phi ']} = \int D \phi \ K[\phi ' ,\phi ] e^{-S[ \phi ]}.
\end{equation}
The kernel $K[\phi ',\phi ]$ has to be chosen such that the partition
function and all expectation values remain invariant under the RGT. At
the end, one usually rescales the lattice spacing back to 1.  In any
case, the correlation length in lattice units gets divided by $n$.

For the kernel functional there are many possible choices \cite{BA92}.
A particularly simple choice for the kernel functional is
\begin{equation} \label{deltargt}
K[ \phi ' , \phi ] = \prod_{x'} \delta \Big( \phi'_{x'} - 
\frac{\beta_{n}}{n^{d}} \sum_{x \in x'} \phi _{x} \Big) \ .
\end{equation}
Assume that we are on a ``critical surface'', where the correlation
length is infinite. With a suitably chosen renormalization factor
$\beta_{n}$ we obtain---after an infinite number of RGT iterations---a
finite {\em fixed point action} (FPA) $S^{*}[\phi ]$. An FPA is
invariant under the RGT.  The task of $\beta_{n}$ is the
neutralization of the rescaling of the field $\phi '$ at the end.
The FPA is an example of a {\em perfect lattice action} since it
is insensitive to a change of the lattice spacing.

Eq. (\ref{deltargt}) can be generalized, for instance, to a Gaussian
form of blocking.  For free fermions, a generalization of this type
reads (we ignore constant factors in the partition function)
\begin{eqnarray}
\lefteqn{e^{-S'[\bar \Psi ' , \Psi ' ]} 
= \int D \bar \Psi D \Psi
e^{-S[\bar \Psi , \Psi ]}} \label{trafo}\nonumber \\
&\times& \exp \Big\{ - \frac{1}{\alpha} \sum_{x'} \Big[
\bar \Psi^{'}_{x'} - \frac{1}{n^{(d+1)/2}} \sum_{x\in x'} \bar \Psi_{x} \Big]
\Big[ \Psi^{'}_{x'} - \frac{1}{n^{(d+1)/2}} \sum_{x\in x'} \Psi_{x} \Big]
\Big\} .
\end{eqnarray}
Here we have already inserted the suitable parameter $\beta_{n}$ and
we introduce a new RGT parameter $\alpha$, which is arbitrary.  The
critical surface requires a fermion mass $m=0$, but we can generalize
the consideration to a finite mass.

Assume that we want to perform a number $N$ of RGT
iterations. If we start from a small mass $m/(nN)$,
then the final mass will be $m$. In the limit $nN \to \infty$
we obtain a perfect action at finite mass.
In this context, ``perfect'' means that scaling quantities
do not depend on the lattice spacing, hence they are identical
to the continuum values.

For the above transformation (\ref{trafo}), this perfect action
can be computed analytically in momentum space \cite{BW96}.
The computation simplifies if we let $n\to \infty$, so
that $N=1$ is sufficient. Hence we start from the continuum
action now, and the perfect action takes the form
\begin{eqnarray}
S^{*}[\bar \Psi , \Psi ] &=& \frac{1}{(2\pi )^{d}}
\int_{-\pi}^{\pi} d^{d}p \ \bar \Psi (-p) \Delta^{*} (p)^{-1} \Psi (p) 
\nonumber \\
\Delta^{*}(p) &=& \sum_{l \in \Z^{d}} \frac{\Pi (p+ 2\pi l)^{2}}
{i (p_{\mu}+2\pi l_{\mu}) \gamma_{\mu}+m} + \alpha \ , \quad
\Pi (p) = \prod_{\mu =1}^{d} \frac{2 \sin (p_{\mu}/2)}{p_{\mu}},
\end{eqnarray}
where $\Delta^{*}$ is the perfect propagator. The same perfect action
is obtained starting from a variety of lattice actions, in particular
from the Wilson fermion action.

In coordinate space we write this action as
\begin{equation}
S^{*}[\bar \Psi , \Psi ] = \sum_{x,r} \bar \Psi_{x}
[ \rho_{\mu}(r)\gamma_{\mu} + \lambda (r) ] \Psi_{x+r} .
\label{lam}
\end{equation}
For $\alpha > 0$, where the RGT breaks the chiral symmetry
explicitly, the couplings in $\rho_{\mu}$ and $\lambda$ 
decay exponentially as $\vert r \vert  $ increases.
An exception is the case $d=1$, where they are confined to one
lattice spacing for the special choice
\begin{equation}
\alpha = \frac{e^{m}-m-1}{m^{2}}\, .
\end{equation}
It turns out that for this choice of $\alpha$ the locality is also
excellent in higher dimensions, \ie, the exponential decay of the
couplings is very fast.  This is important, because for practical
purposes the couplings have to be truncated to a short range, and the
truncation should not distort the perfect properties too much.  An
elegant truncation scheme uses periodic boundary conditions over 3
lattice spacings and thus confines the couplings to a unit hypercube.
This yields the HF, with spectral and thermodynamic properties still
drastically improved compared to Wilson fermions \cite{BIE96,BIE98}.

\begin{figure}[htb]
\centerline{\includegraphics[width=.5\textwidth]{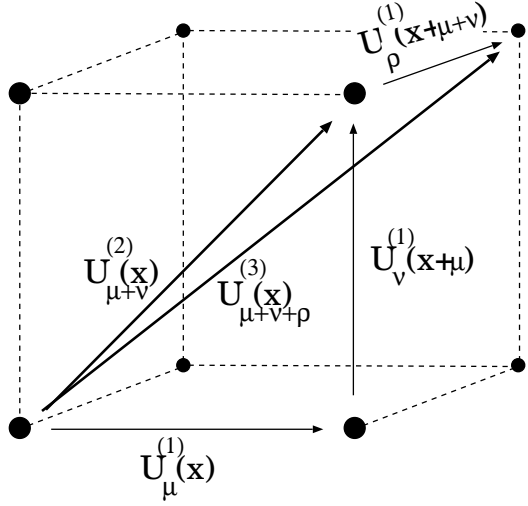}}
\caption{1-space, 2-space and 3-space hyper-links.
\label{links}}
\end{figure}

Of course, it is far more difficult to construct an approximately
perfect action for a complicated interacting theory like QCD.
However, as a simple ansatz we can just use HF together with the
standard gauge link variables.  Apart from nearest neighbors, we also
have couplings over 2, 3 and 4-space diagonals in the unit hypercube.
We connect all these coupled sites by all possible {\em shortest}
lattice paths, by multiplying the compact gauge fields on the path
links.  We call this procedure ``gauging the HF by hand''.  Note that
one can connect two sites $x$ and $y$ lying on 2, 3, and 4-space
diagonals via $d!$ such shortest lattice paths.  We average over all
of them to construct the hyper-link, see \fig{links}.

Let us identify the hyper-link $U^{(1)}_{\mu}(x)$ between site $x$ and
$x+\hat{\mu}$ with $U_{\mu}(x)$, and let us denote the hyper-link in
plane, cube and hyper-cube as $U^{(2)}_{\mu+\nu}(x)$,
$U^{(3)}_{\mu+\nu+\rho}(x)$, and $U^{(4)}_{\mu+\nu+\rho+\sigma}(x)$,
respectively.  Then we can write the corresponding fermion matrix in
terms of the hyper-links which can be constructed recursively starting
from the gauge links $U^{(1)}_{\mu}$,
\begin{eqnarray}
U^{(d)}_{\mu_1+\mu_2+\dots+\mu_d}(x)
=&\frac{1}{d}\Big[&
U^{(1)}_{\mu_1}(x)\,
U^{(d-1)}_{\mu_2+\mu_3+\dots+\mu_d}(x+\hat{\mu}_1) \nonumber\\
&+&U^{(1)}_{\mu_2}(x)\,
U^{(d-1)}_{\mu_1+\mu_3+\dots+\mu_d}(x+\hat{\mu}_2) \nonumber\\
&+&\dots\nonumber\\
&+&U^{(1)}_{\mu_d}(x)\,
U^{(d-1)}_{\mu_1+\mu_2+\dots+\mu_{d-1}}(x+\hat{\mu}_d)\;\;\Big].
\end{eqnarray}

It is convenient to introduce pre-factors which are functions of the
HF hopping parameters $\kappa_i$ and $\lambda_i$, $i=1,\dots,4$, and
sums of $\gamma$-matrices:
\begin{eqnarray}
\Gamma_{\pm\mu} &=& \lambda_1 +\kappa_1(\pm\gamma_{\mu})\nn\\
\Gamma_{\pm\mu\pm\nu} &=& \lambda_2 +\kappa_2(\pm\gamma_{\mu}\pm\gamma_{\nu})\nn\\
\Gamma_{\pm\mu\pm\nu\pm\rho} &=& \lambda_3 +\kappa_3(\pm\gamma_{\mu}\pm\gamma_{\nu}\pm\gamma_{\rho})\nn\\
\Gamma_{\pm\mu\pm\nu\pm\rho\pm\sigma} &=& \lambda_4
+\kappa_4(\pm\gamma_{\mu}\pm\gamma_{\nu}\pm\gamma_{\rho}\pm\gamma_{\sigma})\,
.
\label{lamm}
\end{eqnarray}
Note that the $\lambda_i$ in \eq{lamm} differ from $\lambda(r)$ in
\eq{lam} by a normalization factor $\frac{1}{\lambda_0}$. The
$\kappa_i$ arise from $\rho_{\mu}(r)$ by the same normalization.

\renewcommand{\arraystretch}{1}
\newcommand{\pls}{{\!\!\!+\!}}
\newcommand{\summm}{\!\sum}
The corresponding HF matrix is defined by
\begin{equation}\scriptscriptstyle
\begin{array}{rlllllrrrr}
M_{HF}(x,y)& = \lambda_{0}\Bigg\{ \delta_{x,y} \\
+\summm\limits_{\mu}\Big[\\
\Gamma_{+\mu}
&U^{(1)}_\mu(x)         \delta_{x,y-\hat \mu}&\pls&
\Gamma_{-\mu}
U^{(1)}_{-\mu}(x)       \delta_{x,y+\hat\mu}\;\;\Big]\\
\\
+\summm\limits_{{\mu}}\summm\limits_{{\nu>\mu}}\Big[\\
\Gamma_{+\mu+\nu}
&
U^{(2)}_{+\mu+\nu}(x)   \delta_{x,y-\hat\mu-\hat\nu}&\pls&
\Gamma_{+\mu-\nu}
U^{(2)}_{+\mu-\nu}(x)   \delta_{x,y-\hat\mu+\hat\nu}\;\;\\
                              +
\Gamma_{-\mu+\nu}
&
U^{(2)}_{-\mu+\nu}(x)   \delta_{x,y+\hat\mu-\hat\nu}&\pls&
\Gamma_{-\mu-\nu}
U^{(2)}_{-\mu-\nu}(x)   \delta_{x,y+\hat\mu+\hat\nu}\;\;\Big]\\
\\
+\summm\limits_{{\mu}}
\summm\limits_{{\nu>\mu}}
\summm\limits_{{\rho>\nu}}\Big[\\
\Gamma_{+\mu+\nu+\rho}
&
U^{(3)}_{\mu+\nu+\rho}(x)
\delta_{x,y-\hat\mu-\hat\nu-\hat\rho}&\pls&
\Gamma_{+\mu+\nu-\rho}
U^{(3)}_{\mu+\nu-\rho}(x)   \delta_{x,y-\hat\mu-\hat\nu+\hat\rho}\\
+
\Gamma_{+\mu-\nu+\rho}
&
U^{(3)}_{\mu-\nu+\rho}(x)   \delta_{x,y-\hat\mu+\hat\nu-\hat\rho}&\pls&
\Gamma_{+\mu-\nu-\rho}
U^{(3)}_{\mu-\nu-\rho}(x)   \delta_{x,y-\hat\mu+\hat\nu+\hat\rho}\;\;\\
+
\Gamma_{-\mu+\nu+\rho}
&
U^{(3)}_{-\mu+\nu+\rho}(x)
\delta_{x,y+\hat\mu-\hat\nu-\hat\rho}&\pls&
\Gamma_{-\mu+\nu-\rho}
U^{(3)}_{-\mu+\nu-\rho}(x)   \delta_{x,y+\hat\mu-\hat\nu+\hat\rho}\\
+
\Gamma_{-\mu-\nu-\rho}
&
U^{(3)}_{-\mu-\nu+\rho}(x)
\delta_{x,y+\hat\mu+\hat\nu-\hat\rho}&\pls&
\Gamma_{-\mu-\nu-\rho}
U^{(3)}_{-\mu-\nu-\rho}(x)   \delta_{x,y+\hat\mu+\hat\nu+\hat\rho}\;\;\Big]\\
\\
+\summm\limits_{{\mu}}
\summm\limits_{{\nu>\mu}}
\summm\limits_{{\rho>\nu}}
\summm\limits_{{\sigma>\rho}}\Big[\\
\Gamma_{+\mu+\nu+\rho+\sigma}&
U^{(4)}_{\mu+\nu+\rho+\sigma}(x)\delta_{x,y-\hat\mu-\hat\nu-\hat\rho-\hat\sigma}&\pls&
\Gamma_{+\mu+\nu+\rho-\sigma}
U^{(4)}_{\mu+\nu+\rho-\sigma}(x)\delta_{x,y-\hat\mu-\hat\nu-\hat\rho+\hat\sigma}\\
+\Gamma_{+\mu+\nu-\rho+\sigma}&
U^{(4)}_{\mu+\nu-\rho+\sigma}(x)\delta_{x,y-\hat\mu-\hat\nu+\hat\rho-\hat\sigma}&\pls&
\Gamma_{+\mu+\nu-\rho-\sigma}
U^{(4)}_{\mu+\nu-\rho-\sigma}(x)\delta_{x,y-\hat\mu-\hat\nu+\hat\rho+\hat\sigma}\\
+\Gamma_{+\mu-\nu+\rho+\sigma}&
U^{(4)}_{\mu-\nu+\rho+\sigma}(x)\delta_{x,y-\hat\mu+\hat\nu-\hat\rho-\hat\sigma}&\pls&
\Gamma_{+\mu-\nu+\rho-\sigma}
U^{(4)}_{\mu-\nu+\rho-\sigma}(x)\delta_{x,y-\hat\mu+\hat\nu-\hat\rho+\hat\sigma}\\
+\Gamma_{+\mu-\nu-\rho+\sigma}&
U^{(4)}_{\mu-\nu-\rho+\sigma}(x)\delta_{x,y-\hat\mu+\hat\nu+\hat\rho-\hat\sigma}&\pls&
\Gamma_{+\mu-\nu-\rho-\sigma}
U^{(4)}_{\mu-\nu-\rho-\sigma}(x)\delta_{x,y-\hat\mu+\hat\nu+\hat\rho+\hat\sigma}\;\;\\
+
\Gamma_{-\mu+\nu+\rho+\sigma}&
U^{(4)}_{-\mu+\nu+\rho+\sigma}(x)\delta_{x,y+\hat\mu-\hat\nu-\hat\rho-\hat\sigma}&\pls&
\Gamma_{-\mu+\nu+\rho-\sigma}
U^{(4)}_{-\mu+\nu+\rho-\sigma}(x)\delta_{x,y+\hat\mu-\hat\nu-\hat\rho+\hat\sigma}\\
+\Gamma_{-\mu+\nu-\rho+\sigma}&
U^{(4)}_{-\mu+\nu-\rho+\sigma}(x)\delta_{x,y+\hat\mu-\hat\nu+\hat\rho-\hat\sigma}&\pls&
\Gamma_{-\mu+\nu-\rho-\sigma}
U^{(4)}_{-\mu+\nu-\rho-\sigma}(x)\delta_{x,y+\hat\mu-\hat\nu+\hat\rho+\hat\sigma}\\
+\Gamma_{-\mu-\nu+\rho+\sigma}&
U^{(4)}_{-\mu-\nu+\rho+\sigma}(x)\delta_{x,y+\hat\mu+\hat\nu-\hat\rho-\hat\sigma}&\pls&
\Gamma_{-\mu-\nu+\rho-\sigma}
U^{(4)}_{-\mu-\nu+\rho-\sigma}(x)\delta_{x,y+\hat\mu+\hat\nu-\hat\rho+\hat\sigma}\\[0pt]
+\Gamma_{-\mu-\nu-\rho+\sigma}&
U^{(4)}_{-\mu-\nu-\rho+\sigma}(x)\delta_{x,y+\hat\mu+\hat\nu+\hat\rho-\hat\sigma}&\pls&
\Gamma_{-\mu-\nu-\rho-\sigma}
U^{(4)}_{-\mu-\nu-\rho-\sigma}(x)\delta_{x,y+\hat\mu+\hat\nu+\hat\rho+\hat\sigma}
\;\;\\
&\Big]\Bigg\}.
&&\qquad
\qquad
\qquad
\qquad
\qquad
\qquad
\qquad
(\ref{HFMATRIX})
\end{array}\nonumber\displaystyle
\label{HFMATRIX}
\end{equation}

The sums in \eq{HFMATRIX} run over four different directions for two
1-space links, six directions for four 2-space links, four directions
for eight 3-space links, and one direction for the sixteen 4-space
links. Altogether 80 hyper-links contribute.

With each path of the free HF, a $\gamma$-matrix is associated.  We
have chosen the $\gamma$-matrices so that they add up to produce an
effective $\Gamma$, see \eq{lamm}, which is associated with a given
hyper-link.

The 1-space links $U^{(1)}_{\mu}(x)$ are identical to the
hermitean-conjugate links in negative direction,
$U^{\dagger(1)}_{-\mu}(x+\hat\mu)$, and this feature also holds for
the hyper-links, \eg,
\begin{equation}
U^{\dagger(d)}_{\mu_1+\mu_2+\dots+\mu_d}(x)=
U^{(d)}_{-\mu_1-\mu_2-\dots-\mu_d}(x+\hat\mu_1+\hat\mu_2+\dots+\hat\mu_d).
\end{equation}
Therefore, only one half of the hyper-links have to be stored in
the implementation of the HF.

As in case of Wilson fermions, the HF matrix exhibits the 
``$\gamma_5$ symmetry'', 
\begin{equation}
\gamma_5\,M_{HF}\gamma_5=M_{HF}^{\dagger},
\end{equation}
\ie, $M_{HF}$ is non-hermitean but its eigenvalues come in
complex-conjugate pairs\footnote{Using such a fermionic action at
  $m=0$, one obtains for instance a strongly improved meson dispersion
  relation \cite{BIE96}.  For the inclusion of a truncated perfect
  quark gluon vertex function, see Ref.~\cite{OR97}.}.

\begin{figure}[hbt]
\centerline{\includegraphics[angle=270,width=.48\textwidth]{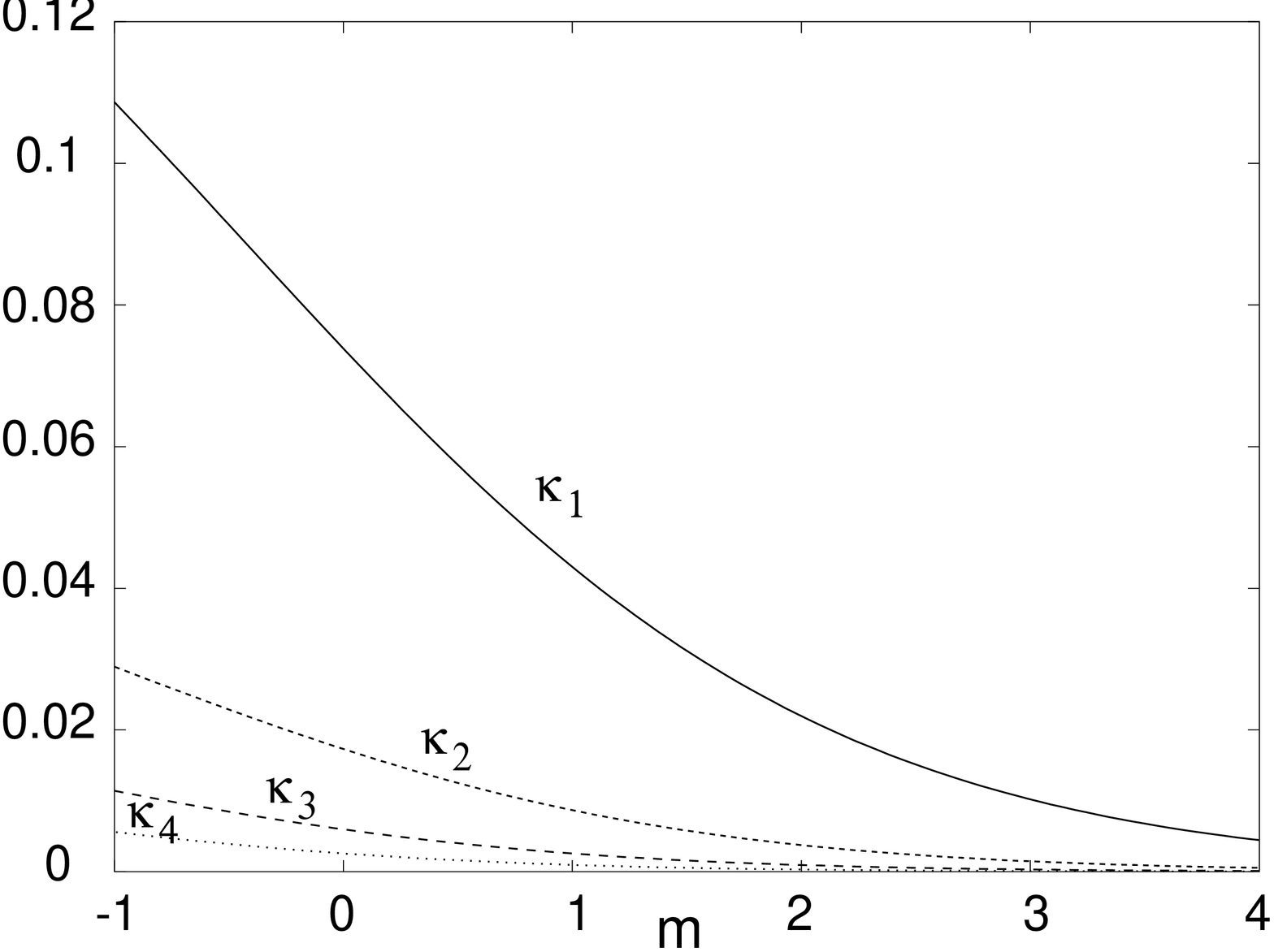}\hfill
\includegraphics[angle=270,width=.48\textwidth]{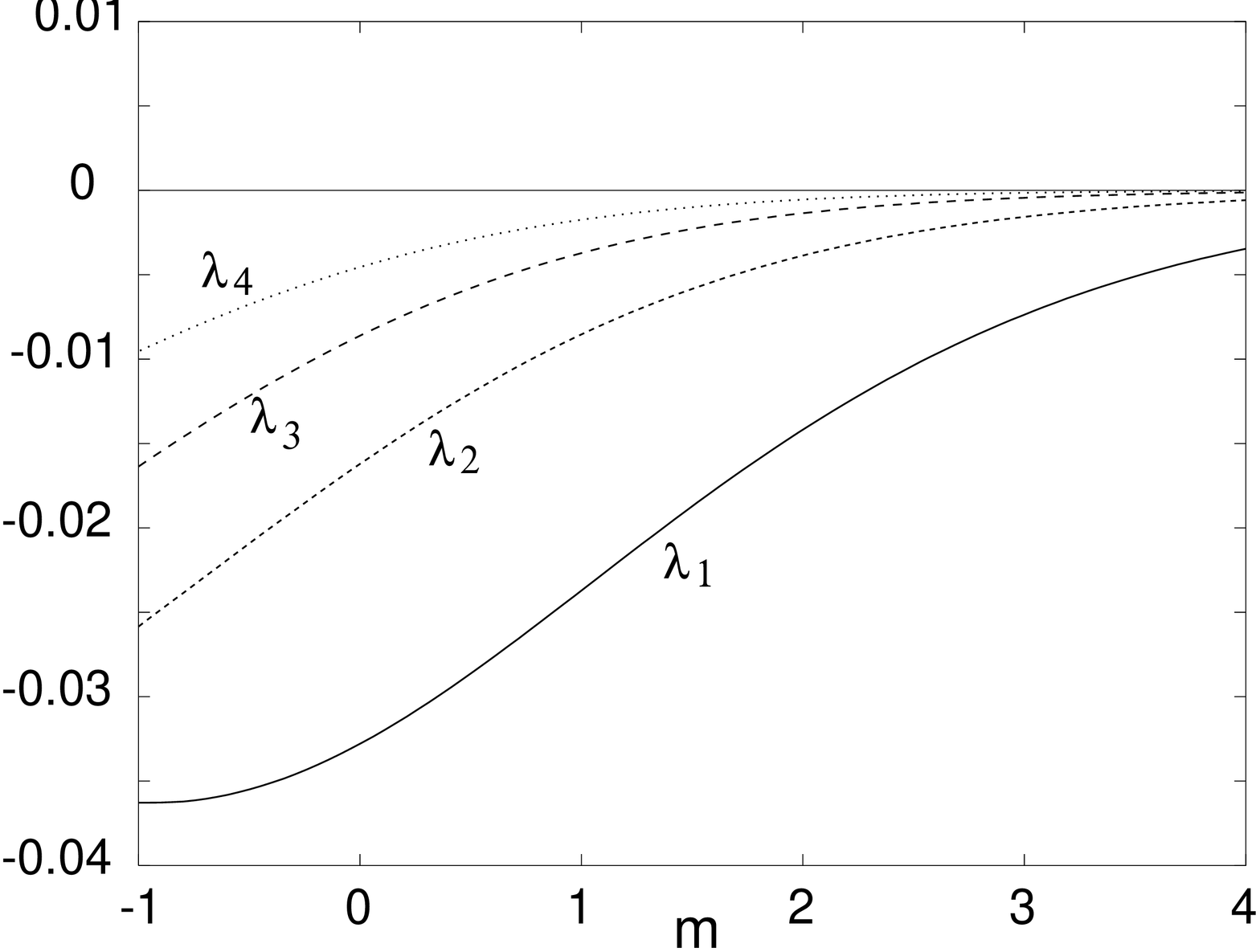}}
\caption{
  The vector ``hopping parameters'' $\kappa_i$\label{hopkin}, and the
  scalar ``hopping parameters'' $\lambda_i$ as functions of the mass
  $m$.}
\end{figure}

The off-diagonal elements (``hopping parameters'') $\kappa_{i}$ and
$\lambda_{i}$ are shown as functions of the mass $m$ in
Fig.~\ref{hopkin}. 


\section{Block SSOR Preconditioning\label{SSOR}}

Preconditioning a linear system 
\begin{equation} \label{Matrix_eq}
M \psi = \phi
\end{equation}
amounts to selecting two regular matrices $V_1$ and $V_2$, which act
as a left and a right preconditioner, respectively. This means that we
consider the modified system\footnote{We mark preconditioned
  quantities by a tilde.}
\begin{equation}
  V_1^{-1}MV_2^{-1} \tilde{\psi} = \tilde{\phi}, \enspace \mbox{where
    } \tilde{\phi} := V_1^{-1}\phi, \; \tilde{\psi} := V_2\psi.
  \label{Matrix_prec_eq}
\end{equation}
The spectral properties of the preconditioned matrix
$V_{1}^{-1}MV_{2}^{-1}$ depend only on the product $V_1V_2$, but not
on the manner how it is factorised into $V_1$ and $V_2$.  For a good
choice, the number of iteration steps required for solving
eq.~\eqnref{Matrix_prec_eq} by Krylov subspace methods (such as
BiCGStab) can be reduced significantly, compared to the original
system (\ref{Matrix_eq}).

In this paper, we consider block SSOR preconditioning.  Let $M = D - L
- U$ be the decomposition of $M$ into its block diagonal part $D$, its
(block) lower triangular part $-L$ and its (block) upper triangular
part $-U$.  Given a relaxation parameter $\omega \not = 0$, block SSOR
preconditioning is then defined through the choice
\begin{equation}
V_1 = \left( \frac{1}{\omega}D - L \right) \left( \frac{1}{\omega}D 
\right)^{-1},
 \enspace V_2 = \frac{1}{\omega}D - U \; .
\label{DEFOMEGA}
\end{equation}
It is important that block SSOR preconditioning is particularly cheap
in terms of arithmetic costs due to the {\em Eisenstat trick}
\cite{Eis}, which is based on the identity
\begin{eqnarray}
\lefteqn{ 
\frac{1}{\omega}D \left(\frac{1}{\omega}D - L \right) ^{-1}
(D - L - U) \left( \frac{1}{\omega}D-U\right)^{-1} 
} \\
&=& \nonumber
\frac{1}{\omega}D \left\{ \left( \frac{1}{\omega}D - L \right)^{-1}
 \left[ \uno + \left( \frac{\omega - 2}{\omega}D \right) 
\left(\frac{1}{\omega}D - U
\right)^{-1} \right] + \left( \frac{1}{\omega}D - U \right)^{-1} \right\} \\
&=&   \left( \uno - \omega LD^{-1} \right)^{-1}  \nonumber  
\left[ \uno +   (\omega - 2) \left(\uno - \omega UD^{-1}  
\right)^{-1} \right] + \left( \uno - \omega UD^{-1} \right)^{-1} .  
\end{eqnarray}
 
Therefore, the matrix-vector product with the preconditioned matrix,
$z = V^{-1}_1MV^{-1}_2 x$, can be computed according to Algorithm~1.

\begin{center}
\begin{minipage}{\textwidth}
\begin{tabbing}
\hspace*{2ex} \= \hspace{2ex} \=  \hspace{2ex} \= \hspace{2ex} \kill
\> solve $(\uno - \omega UD^{-1}) y = x$ \\
\> compute $w = x + (\omega - 2) y $ \\
\> solve $(\uno - \omega LD^{-1}) v = w$ \\
\> compute $z = v + y$
\end{tabbing}  
\end{minipage}

{Alg. 1} 
Matrix-vector product for the preconditioned system.
\end{center}

Here the matrices $\uno - \omega U D^{-1}$ and $\uno - \omega L
D^{-1}$ are (non-block) lower and upper triangular, respectively. This
means that solving the corresponding linear systems amounts to a
simple forward and backward substitution process.  Algorithm~2 gives a
detailed description of the forward substitution for solving $(\uno -
\omega LD^{-1})v = w$. Here we denote the block components of the
vectors $v,w$ and the matrices $D,L$ as $v_i, w_i$ and $D_{ii}$,
$L_{ij}$.  The backward substitution for $(\uno - \omega U D^{-1}) y
=x$ can be carried out analogously.

\begin{center}
\begin{minipage}{\textwidth}
\begin{tabbing}
\hspace*{2ex} \= \hspace{2ex} \=  \hspace{2ex} \= \hspace{2ex} \kill
\> for $i=1$ to $N$ \\
\> \> $v_i = w_i + \sum_{j=1}^{i-1} L_{ij} s_j$  \\
\> \> $s_i = \omega D_{ii}^{-1}v_i$ 
\end{tabbing} 
\end{minipage}

{Alg. 2} 
Forward substitution.
\end{center}

Assuming that the blocks $D_{ii}$ of $D$ are dense and that their
inverses have been pre-computed, we see that one iteration step in the
above algorithm is exactly as expensive as a direct multiplication
with the matrix $D - L$ (except for the additional multiplication with
$\omega$). A similar relation holds for the backward substitution and
the multiplication with $D-U$.  Note that the two multiplications with
$D-L$ and $D-U$ are as expensive as one multiplication with the whole
matrix $M$ plus one additional multiplication with $D$. This allows us
to quantify exactly the work required when using block SSOR
preconditioning with the Eisenstat trick.

\begin{itemize}
\item {\em Initialization}: the inverses $D^{-1}_{ii}$ of all diagonal
  blocks of the block diagonal matrix $D$ must be pre-computed before
  the iteration starts. We also assume that these inverses are already
  scaled by the factor $\omega$ in the initialization.
\item {\em Iteration}: each iterative step requires additional
  arithmetic work equivalent to one matrix-vector multiplication with
  the matrix $D$ plus one vector scaling (with factor
  $\omega-2$) and two vector summations.
\end{itemize} 

In SU(3) lattice gauge theory, a natural choice for the block
diagonal matrix $D$ takes each block $D_{ii}$ to be of dimension
12, corresponding to the 12 variables residing at each lattice point.
In this work, we will consider this choice, denoted as $D^{(12)}$, as
well as the three generic options $D^{(6)}$, $D^{(3)}$ and $D^{(1)}$,
where the diagonal blocks are of dimension 6, 3 and 1, respectively.
The choices $D^{(6)}$ and $D^{(3)}$ also appear `natural'---at least
within the SWA framework---since a diagonal block of $D^{(12)}$
carries the structure of eq.~\eqnref{eq:structure}.  Accordingly,
ignoring the parameters $\kappa$ and $c_{SW}$, four consecutive $3
\times 3$ blocks in $D^{(3)}$ are given by $\one + F_1$, $\one - F_1$,
$\one + F_1$, $\one - F_1$ and two consecutive $6 \times 6$ diagonal
blocks in $D^{(6)}$ are identical and given by
\begin{equation}
\left( 
    \begin{array}{cc} 
      {\mathbf 1}+F_1&F_2\\ 
      F_2^\dagger&{\mathbf 1}-F_1 
    \end{array}  
  \right) \; .
\end{equation}

Table~\ref{cost_table} quantifies the arithmetic effort for computing
a matrix-vector product with each of the matrices
 $D^{(12)}$, $D^{(6)}$,
$D^{(3)}$ and $D^{(1)}$. We count this effort in units of
{\em cflops}, which represent one multiplication of complex numbers 
followed by one summation. The table also quotes estimates
for the arithmetic work to compute the inverse of each of these
matrices in units of matrix-vector multiplies (MVM). Precise numbers
will depend on the particular algorithm chosen for the inversion. 
The estimates in Table~\ref{cost_table} are based on a particularly
efficient way for computing the inverse, which uses Cramer's rule on $3\times3$
blocks and which exploits the additional block structure within each of 
the $D^{(6)}_{ii}$ and $D^{(12)}_{ii}$.   

The percentages given in brackets quantify these numbers in terms of
the cost for a single matrix-vector multiply with $M$.  Referring to
our previous discussion, they specify the additional cost for (block)
SSOR preconditioning. $V$ denotes the lattice volume and one
matrix-vector multiplication with $M$ is counted with $528V$ {\em
  cflops}.

\begin{table}
\caption{Arithmetic cost  
for different sizes of the diagonal blocks. 
$V$ is the lattice volume.\label{cost_table}}
\begin{center}
\begin{tabular}{rcccc}
 \hline
  & $D^{(12)}$ & $ D^{(6)}$  & $D^{(3)} $ & $D^{(1)}$  \\ \hline
 MVM (in {\em cflops}) 
& 144$V$ ($27.2\%$) &72$V$ ($13.6\%$) & 36$V$ ($6.8\%$) & $V$ ($0.2\%$) \\  
inverse (in MVM) 
&  10 & 2.5 & 2 & 1  \\ \hline
\end{tabular}
\end{center}
\end{table} 

\section{Parallelization\label{PARA}}

In the fermion equation \eqnref{Matrix_eq}, we have the
freedom to choose any ordering for the lattice points $x$.
Different orderings yield different matrices $M$, which are
permutationally similar to each other. One matrix can be retrieved
from the other one by the transformation $M \rightarrow
P^{\dagger}MP$, where $P$ is a permutation matrix.  In general,
the quality of the block SSOR preconditioner depends
on the ordering scheme.

On the other hand, the ordering of the lattice points also determines
the degree of parallelism within the forward (and backward)
substitutions as described in Algorithm~3. Usually, there is a
trade-off between the parallelism a given ordering allows for, and the
efficiency of the corresponding SSOR preconditioning.

In an earlier paper \cite{FIS96}, we have shown that for the non-block
SSOR preconditioner and the Wilson fermion matrix one can use a {\em
  locally lexicographic} ordering on parallel computers supporting grid
topologies, so that the resulting
SSOR preconditioner parallelizes nicely while significantly
outperforming the standard odd-even preconditioner. The purpose of
this section is to show that this is also possible for the block SSOR
preconditioners considered here, even for situations where $M$
represents couplings beyond nearest-neighbor lattice points.

Let $n(x)$ denote the set of all lattice points a given site $x$ is
coupled to. For example, $n(x) = \{ y \mid y = x \pm \hat\mu,~\mu =
1,\ldots,4 \}$ for the nearest-neighbor coupling, $n(x) = \{ y\ne x \mid
\sum_{\mu = 1}^{4} |x_{\mu} - y_{\mu}| \leq 2 \} $ for nearest and
next-to-nearest neighbor couplings, or $n(x) = \{ y\ne x \mid |x_{\mu} -
y_{\mu}|\leq 1, ~ \mu = 1, \dots ,4\}$ for the hypercube couplings.

We now re-formulate the forward substitution of Algorithm~2 for this
generalized situation.  We assume an overall natural partitioning of
$M$ into sub-blocks of dimension 12 (corresponding to the 12 variables
at a given lattice point), and we use the lattice positions $x,y$ as
indices for those blocks.  By $D$ we denote any of the matrices
$D^{(12)}, \ldots, D^{(1)}$, so $D_{xx}$ stands for a diagonal block
of dimension $12 \times 12$. It is fully occupied in the case $D = D^{(12)}$,
whereas in case $D = D^{(6)}$ it consists of two $6 \times 6$ diagonal
blocks, etc.  We also use the relation $x \leq_{o} y$ between lattice
points to denote that $x$ has been numbered before $y$ with respect to
a given ordering $o$.

\begin{center}
\begin{minipage}{\textwidth}
\begin{tabbing}
\hspace*{2ex} \= \hspace{2ex} \=  \hspace{2ex} \= \hspace{2ex} \kill
\> for all lattice positions $x$ in a given ordering $o$ \\
\> \> $v_x = w_x + \sum_{y \in n(x), \; y \leq_o x } L_{xy} s_y$  \\
\> \> $s_x = \omega D_{xx}^{-1}v_x$ 
\end{tabbing}
\end{minipage}

{Alg. 3} 
Generalized forward substitution.
\end{center}

To discuss parallelization, we use the concept of coloring the lattice
points. A decomposition of all lattice points into mutually disjoint
sets $C_1,\ldots,C_k$ is termed a {\em coloring} (with respect to the
matrix $M$), if for any $l \in \{1,\ldots,k\}$ the property
\[
x \in C_l \Rightarrow y \not \in C_l\; \mbox{ for all }\; y \in n(x)
\]
holds.  Associating a different color with each set $C_l$, this
property means that each lattice point couples with lattice points of
different colors only.  An associated color ordering first numbers all
grid points with color $C_1$, then all with $C_2$ etc. With such a
color ordering we see that the computation of $v_x$ for all $x$ of a
given color $C_l$ can be done in parallel, since terms like $\sum_{y
  \in n(x), \; y \leq_o x }$ involve only lattice points from the
preceding colors $C_1,\ldots,C_{l-1}$.

In the case of nearest-neighbor couplings, the familiar odd-even
ordering represents such a coloring with two colors corresponding to
the odd and the even sublattice.  For the case of the Wilson fermion
matrix, we pointed out in \cite{FIS96} that the (non-block) SSOR
preconditioned system may be interpreted as a representation of the
familiar odd-even reduction process.  A similar relation arises in the
case of  SWA, where the reduced system considered in Ref. \cite{JAN96}
is equivalent to the ($12 \times 12$ block) SSOR preconditioned matrix
with odd-even ordering.

For more complicated couplings like the next-to-nearest neighbor 
couplings or the HF, it would become
increasingly difficult to
handle colorings with a minimum number of colors. For
example, the hypercube ordering requires at least 16 different
colors.

However, aiming at a minimal number of colors is not a good
strategy. For example, the
odd-even coloring can actually be considered as the `worst case' as far
as the quality of the corresponding SSOR preconditioner is concerned
\cite{FIS96}.
Heuristically, this can be explained as follows: if the number of
colors is small, the color sets $C_l$ themselves are large, and
information is not spread between lattice points of
equal color in a forward (or backward) substitution.

Therefore, the right strategy is to search for colorings such that the
number $k$ of colors is {\em maximal}, while the number of points
within each color is still in agreement with the parallelization
we are aiming for.

The {\em locally lexicographic ordering}, proposed in Ref.\ 
\cite{FIS96} for the case of a nearest-neighbor coupling, turns out to
be an adequate ordering also for more complicated couplings like
next-to-nearest neighbor and hypercube.
  
To describe this ordering, we assume the processors of the parallel computer to be
connected as a 4-dimensional grid $p_1 \times p_2 \times p_3 \times p_4$. 
Note that this includes lower dimensional grids by setting some
of the $p_i$ to 1. 
The space-time lattice can be matched to the processor
grid in a natural manner, producing a local lattice of size
$n^{loc}_1 \times n^{loc}_2 \times n^{loc}_3 \times n^{loc}_4$ with
$n^{loc}_i = n_i/p_i$ on each processor. Here we assume for
simplicity that each $p_i$ divides $n_i$, and that we have $n^{loc}_i \geq
2$ for $i = 1,\ldots,4$.

Let us divide the lattice sites into $ n^{loc}$ groups where $n^{loc}
= n^{loc}_1 n^{loc}_2 n^{loc}_3 n^{loc}_4$.  Each group corresponds to
a fixed position within the local grids and contains all grid points
appearing at this position within their respective local grid.
Associating a color with each of the groups, we get a coloring in the
sense of the definition above, as long as the coupling defined through
$M$ is local enough.  More precisely, the sets represent a coloring,
if for all $y \in n(x)$ the relation $|y_{\mu} - x_{\mu}| <
n^{loc}_{\mu}$ holds for $\mu = 1,\ldots,4$. For example, we need
$n_{\mu}^{loc} \geq 2$ for all $\mu$ for the hypercube couplings and
$n_{\mu}^{loc} \geq 3$ for all $\mu$ for the next-to-nearest neighbor
couplings.

A {\em locally lexicographic} ($ll$)
ordering is now defined to be the color
ordering, where all points of a given color are ordered 
after all points with colors, which correspond to lattice positions
on the local grid that are lexicographically preceding
the given color.
In \fig{ll_fig}, this
amounts to the alphabetic ordering of the colors $a$ -- $q$.
This example also illustrates the decoupling obtained through that
ordering for (2-dimensional) nearest-neighbor and hypercube couplings.
\setlength{\unitlength}{1cm}
\begin{figure}
\begin{center}
  \begin{picture}(12,10)
    \thicklines
    \multiput(0,0)(0,1){10}{\multiput(0,0)(1,0){12}{\circle{0.5}}}
    \multiput(1.5,-0.5)(4,0){3}{\line(0,1){10}}
    \multiput(-0.5,2.5)(0,4){2}{\line(1,0){12}}
    \put(2.25,6){\vector(1,0){0.5}}
    \put(5,5.75){\vector(0,-1){0.5}}
    \put(4.25,5){\vector(1,0){0.5}} \put(5.75,5){\vector(-1,0){0.5}}
    \thinlines
    \put(4.25,5.75){\vector(1,-1){0.5}} \put(5.75,5.75){\vector(-1,-1){0.5}}
    \thicklines
    \put(2,4.75){\vector(0,-1){0.5}}
    \thinlines
    \put(1.25,4.75){\vector(1,-1){0.5}} \put(2.75,4.75){\vector(-1,-1){0.5}}
    \thicklines
    \put(4.25,3){\vector(1,0){0.5}} \put(5.75,3){\vector(-1,0){0.5}}
    \put(5,2.25){\vector(0,1){0.5}} \put(5,3.75){\vector(0,-1){0.5}}
    \thinlines
    \put(4.25,3.75){\vector(1,-1){0.5}} \put(5.75,3.75){\vector(-1,-1){0.5}}
    \put(4.25,2.25){\vector(1,1){0.5}} \put(5.75,2.25){\vector(-1,1){0.5}}
    \thicklines
    \put(8,4.75){\vector(0,-1){0.5}}
    \thinlines
    \put(7.25,4.75){\vector(1,-1){0.5}} \put(8.75,4.75){\vector(-1,-1){0.5}}
    \thicklines
    \put(7.25,4){\vector(1,0){0.5}}
    \begin{small}
      \begin{em}
        \multiput(0,1)(0,4){3} { \multiput(0,0)(4,0){3} {
            \put(0,0){\makebox(0,0){g}} \put(1,0){\makebox(0,0){h}}
            \put(2,0){\makebox(0,0){e}} \put(3,0){\makebox(0,0){f}} }
          }
        \multiput(0,0)(0,4){3} { \multiput(0,0)(4,0){3} {
            \put(0,0){\makebox(0,0){l}} \put(1,0){\makebox(0,0){m}}
            \put(2,0){\makebox(0,0){i}} \put(3,0){\makebox(0,0){k}} }
          }
        \multiput(0,3)(0,4){2} { \multiput(0,0)(4,0){3} {
            \put(0,0){\makebox(0,0){p}} \put(1,0){\makebox(0,0){q}}
            \put(2,0){\makebox(0,0){n}} \put(3,0){\makebox(0,0){o}} }
          }
        \multiput(0,2)(0,4){2} { \multiput(0,0)(4,0){3} {
            \put(0,0){\makebox(0,0){c}} \put(1,0){\makebox(0,0){d}}
            \put(2,0){\makebox(0,0){a}} \put(3,0){\makebox(0,0){b}} }
          }
      \end{em}
    \end{small}
  \end{picture}
\end{center}
\vspace*{.3cm}
\caption{Locally lexicographic ordering and forward solve in 2
  dimensions. \label{ll_fig} Nearest-neighbor: straight arrows,
  hyper-cube couplings: straight and thin diagonal arrows.}
\end{figure}
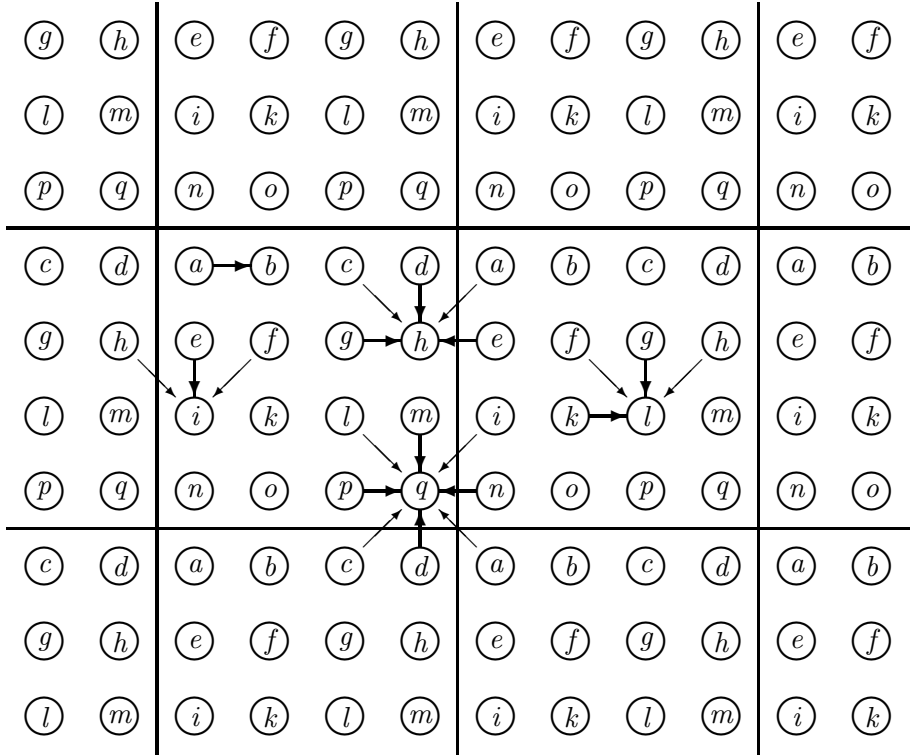

The parallel version of the forward substitution in Algorithm~4 with
the $ll$-ordering now reads:

\begin{center}
\begin{minipage}{\textwidth}
\begin{tabbing}
\hspace*{2ex} \= \hspace{2ex} \=  \hspace{2ex} \= \hspace{2ex} \kill
for all colors in lexicographic order \\
\> for all processors \\
\>  \> $x := $ grid point of the given color on that processor \\
\> \> $v_x = w_x + \sum_{y \in n(x), \; y \leq_{ll} x } L_{xy} s_y$ \\
\> \> $s_x = \omega D_{xx}^{-1}v_x$
\end{tabbing}
\end{minipage}

{Alg. 4} 
$ll$-forward substitution.
\end{center}
 
If the lattice point $x$ is close to the boundary of the local
lattice, then the set $n(x)$ will contain grid points $y$ residing in
neighboring processors.  Therefore, some of the quantities $s_y$ will
have to be communicated from those neighboring processors.  The
detailed communication scheme for the case of a nearest-neighbor
coupling was given in Ref.~\cite{FIS96}. In that case, only the 8 nearest
neighbors in the processor grid were involved in the communication. For
the more complicated HF couplings, all 80 hypercube neighbors may be
involved.  For a 3- or 2-dimensional processor grid, this number
reduces to 26 resp. 8.
  
\section{Results\label{RESULTS}}

The \llssor\ preconditioning of improved actions has been tested in
quenched QCD, for realistic lattice sizes and parameters.  For SWA we
use the odd-even preconditioner of Ref.~\cite{JAN96} as reference.
The HF action preconditioner has been implemented only on a scalar
machine so far.  Work for a parallel implementation is in progress.

\subsection{Sheikholeslami-Wohlert Action}

We are going to compare results from test runs of \llssor\ and
odd-even preconditioning, both codes being equally well optimized for
the multiplication of the SWA fermion matrix.  Our investigations are
based on a de-correlated set of 10 quenched gauge configurations
generated on a $16^4$ lattice at $\beta=6.0$.  We have taken
measurements at 3 values of $c_{SW}$, 0, 1.0 and 1.769.  The latter
value is the optimal quenched $c_{SW}$ coefficient taken from
Ref.~\cite{Lue97}.

In order to provide both machine independent numbers and real time
results on parallel and scalar implementation machines, we will
present iteration numbers which {\em (i)} are directly proportional to
the amount of floating point operations and {\em (ii)} real time
results from implementations on both the parallel system
APE100/Quadrics and a SUN Ultra workstation.

We have applied BiCGStab as iterative solver.  The stopping criterion
has been chosen as $r=\frac{||MX-\phi||}{||X||}\le 10^{-6}$.  We used a
local source.  At the end of the computation, we have checked how far
the true residuum deviates from the accumulated one.  In fact, for
\llssor, the accumulated residuum turned out to deviate only slightly
from the corresponding true residuum.  Moreover, deviations between
the solutions $X$ as computed by \llssor\ and odd-even-preconditioning
have been checked. We found the norms of the solution vectors to
differ in the range of $10^{-6}$.

\begin{figure}%
\begin{center}
\includegraphics[width=.7\textwidth]{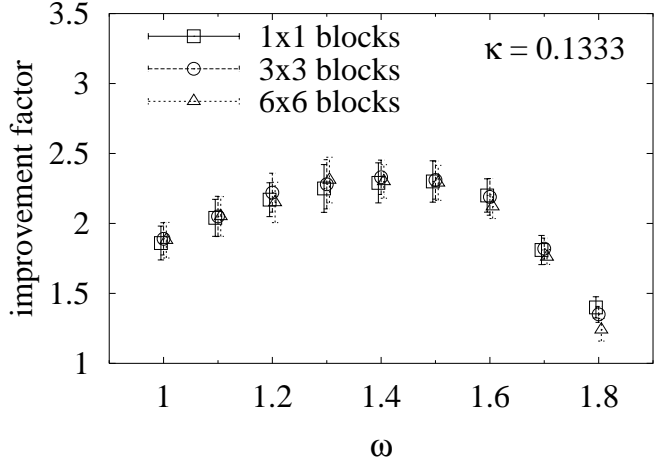}
\end{center}
\caption{
  $\omega$-dependence of \llssor. We plot the gain factor between
  \llssor\ and odd-even preconditioning.
\label{OMEGASWA}}
\end{figure}

In a first step, we have determined the optimal value for the
over-relaxation parameter $\omega$ as introduced in \eq{DEFOMEGA}, see
\fig{OMEGASWA}.  The $\omega$ dependence of the iteration numbers is
measured for $c_{SW}=1.769$ at a given value for the hopping parameter
$\kappa=0.1333$. At this value, the fermion matrix is close to
criticality.

In \fig{OMEGASWA}, the results from three diagonal block sizes are
overlaid, the $1\times 1$, $3\times 3$, and $6\times 6$ blocks.  Only
a weak dependence of the iteration numbers on the block size is
visible, however.  Around $\omega=1.4$ a minimum in iteration numbers
is found\footnote{Ref.~\cite{FIS96} considered only the case
  $\omega=1$ for $c_{SW}=0$.}. We have verified that this number holds
for the whole $\kappa$-range investigated and for the other values of
$c_{SW}$ as well.

\begin{figure}
\begin{center}
{\includegraphics[width=.7\textwidth]{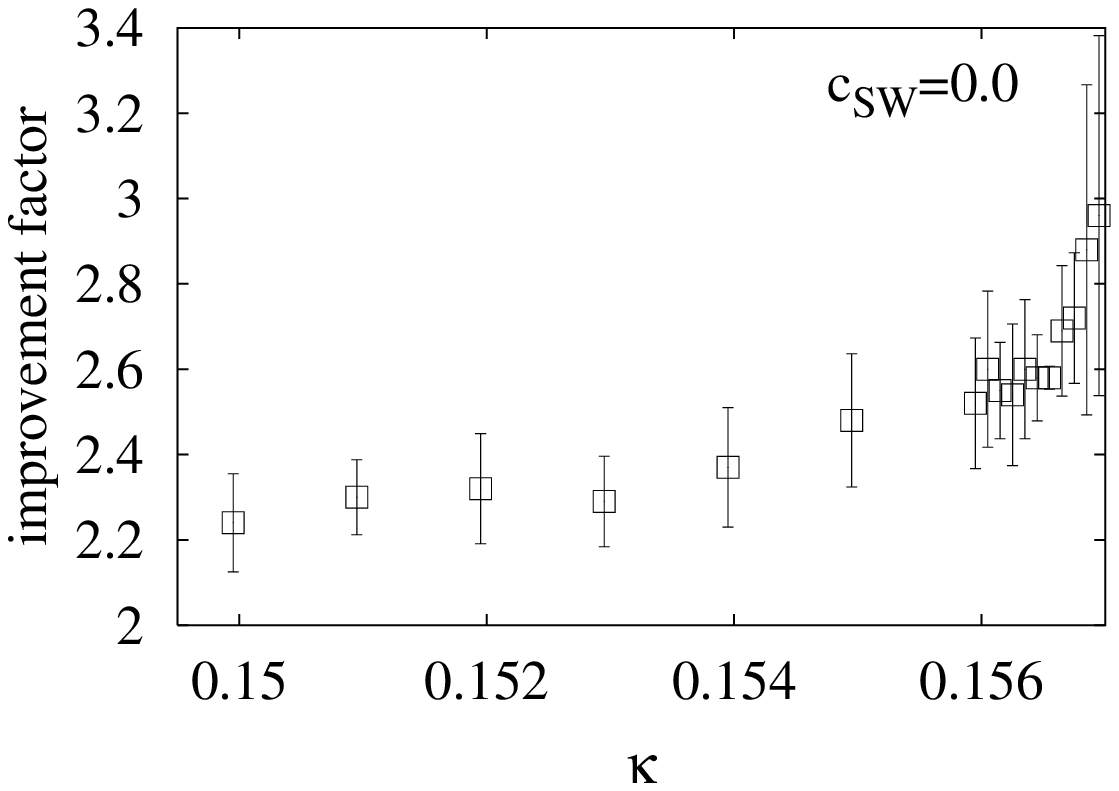}}\\[-.5cm]
{\includegraphics[width=.7\textwidth]{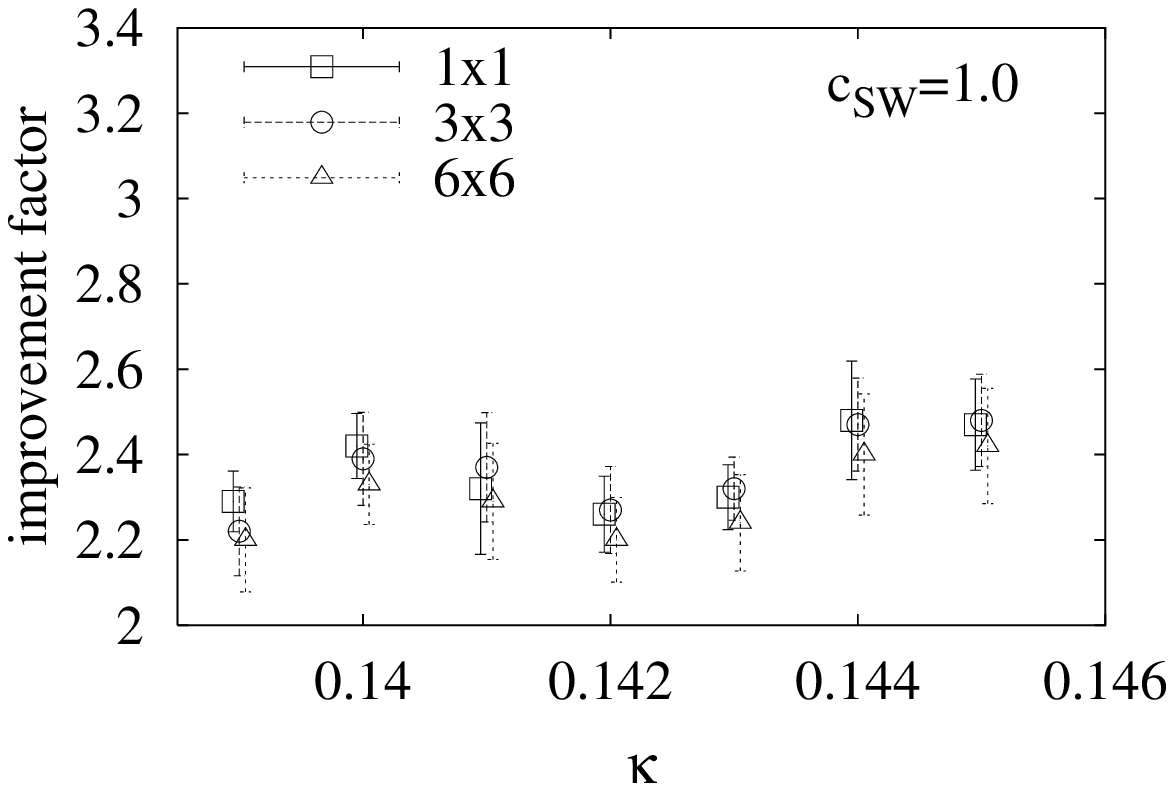}}\\[-.5cm]
{\includegraphics[width=.7\textwidth]{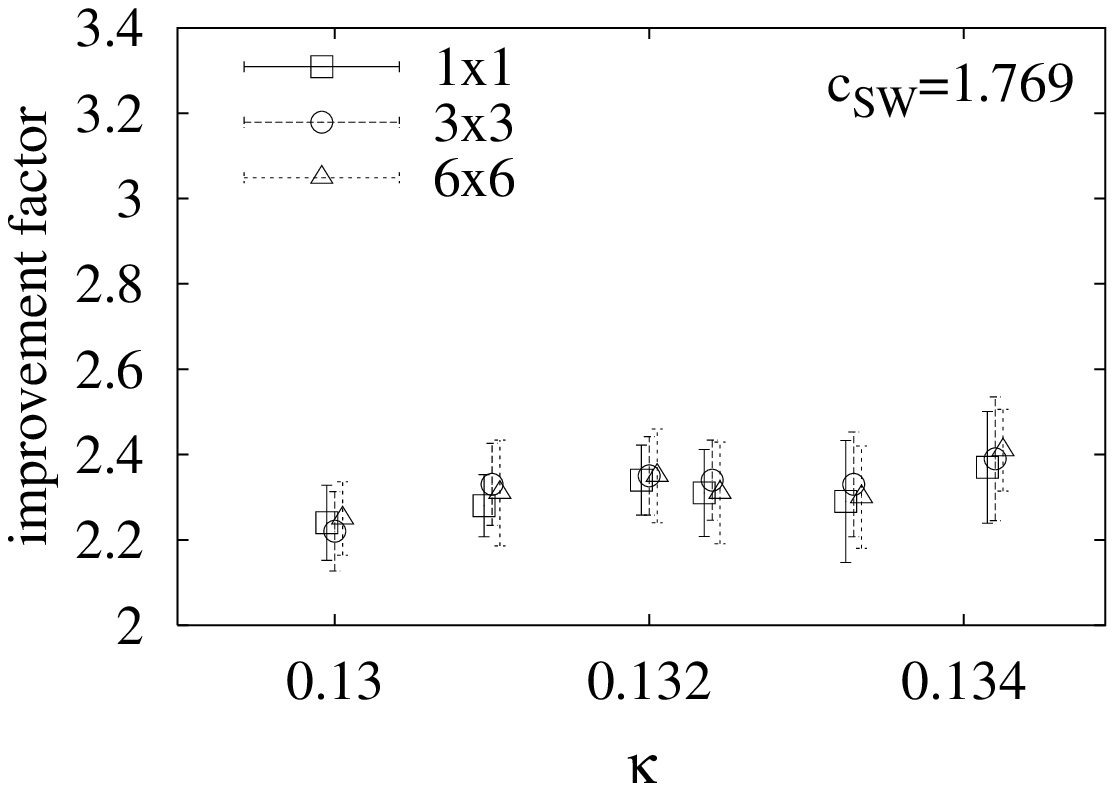}}
\end{center}
\caption{
  $\kappa$ dependence of \llssor\ vs.\ the odd-even preconditioner for
  three values of $c_{SW}$.
\label{OESSOR}}
\end{figure}

Next we benchmark the \llssor\ preconditioner against the odd-even
preconditioner.  In \fig{OESSOR}, the iteration numbers are presented
as a function of $\kappa$, separately for the three values chosen for
$c_{SW}$.  We plot the ratio of iteration numbers of the odd-even
procedure vs.\ \llssor.  In the last two segments of the figure, three
block sizes are overlaid.

The improvement of \llssor\ compared to the odd-even preconditioned
system is rather substantial: close to $m_c$, \ie\ in the region of
interest, a factor up to 2.5 in iteration numbers can be found, with
increasing tendency when approaching $\kappa_c$.  As far as the
dependence of the improvement factor on $c_{SW}$ is concerned, one
cannot detect a systematic effect. Significant block size dependencies
are not visible either.  However, in the actual time measurements on
APE100 to be shown below, we will find the $3\times 3$ local diagonal
block procedure to perform best.

The above results have been achieved on a machine equipped with $p=32$
processors.  With $V=65536$ being the number of sites on the lattice
the sub-lattices comprise $2048$ sites each.  As the regions which are
treated independently in the preconditioning process are as large as
the size of a sub-lattice assigned to a given processor, the
parallelism of the preconditioner follows the number of processors
$p$. The larger the sub-lattices the better is the improvement factor
\cite{FIS96}, since the applicability of the \llssor\ preconditioner
seems to be limited to low granularity.  However, it turns out that on
today's machine sizes---ranging from coarse parallelism with ${\cal
  O}(10)$ processors to massive parallelism with ${\cal O}(1000)$
processors---usual lattice sizes lead to sufficiently large
sub-lattices to ascertain effectively parallel preconditioning.  For
the $3\times 3$ diagonal block \llssor\ procedure, we have
investigated the local lattice size dependence working on four
different APE100/Quadrics systems \cite{QUAD94}, a 32-node Q4, a
128-node QH1, a 256-node QH2, and a 512-node QH4.  For a given lattice
size, the sub-lattice sizes follow the inverse number of nodes.  In
the range investigated, the improvement varies by about 10 \%.

\begin{figure}
  \centerline{\includegraphics[width=.7\textwidth]{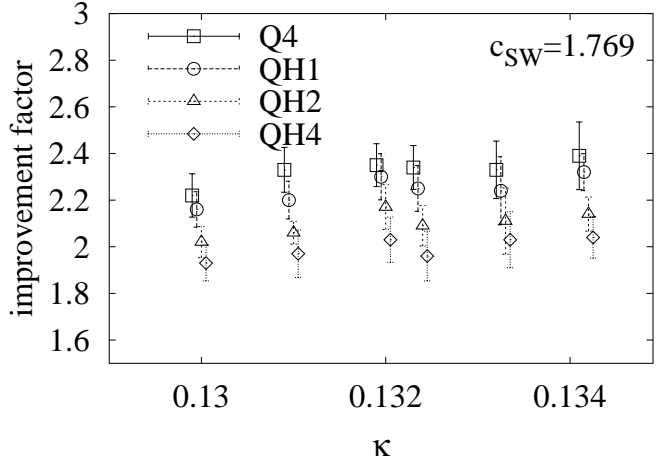}}
\caption{
  Local lattice size dependence of \llssor\ preconditioning as
  determined for a fixed size system on 32, 128, 256 and 512 node
  APE100/Quadrics machines.
\label{LOCAL}}
\end{figure}

\begin{figure}
\begin{center}
{\includegraphics[width=.7\textwidth]{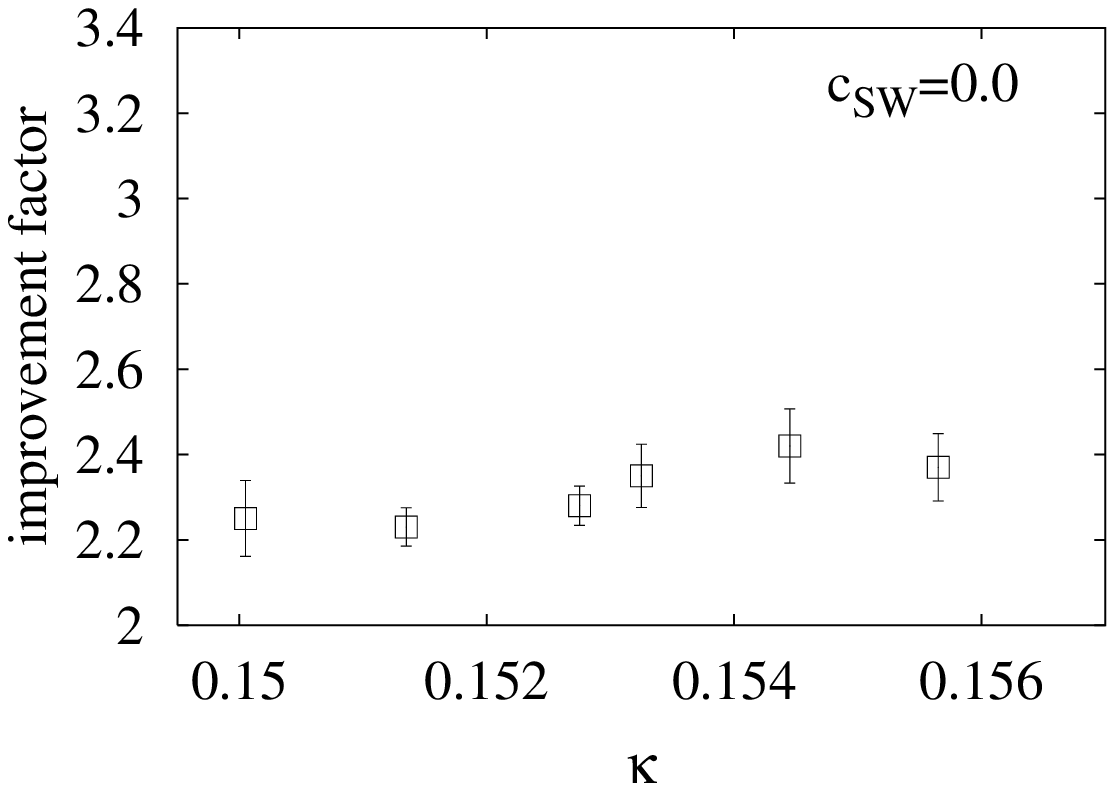}}\\[-.5cm]
{\includegraphics[width=.7\textwidth]{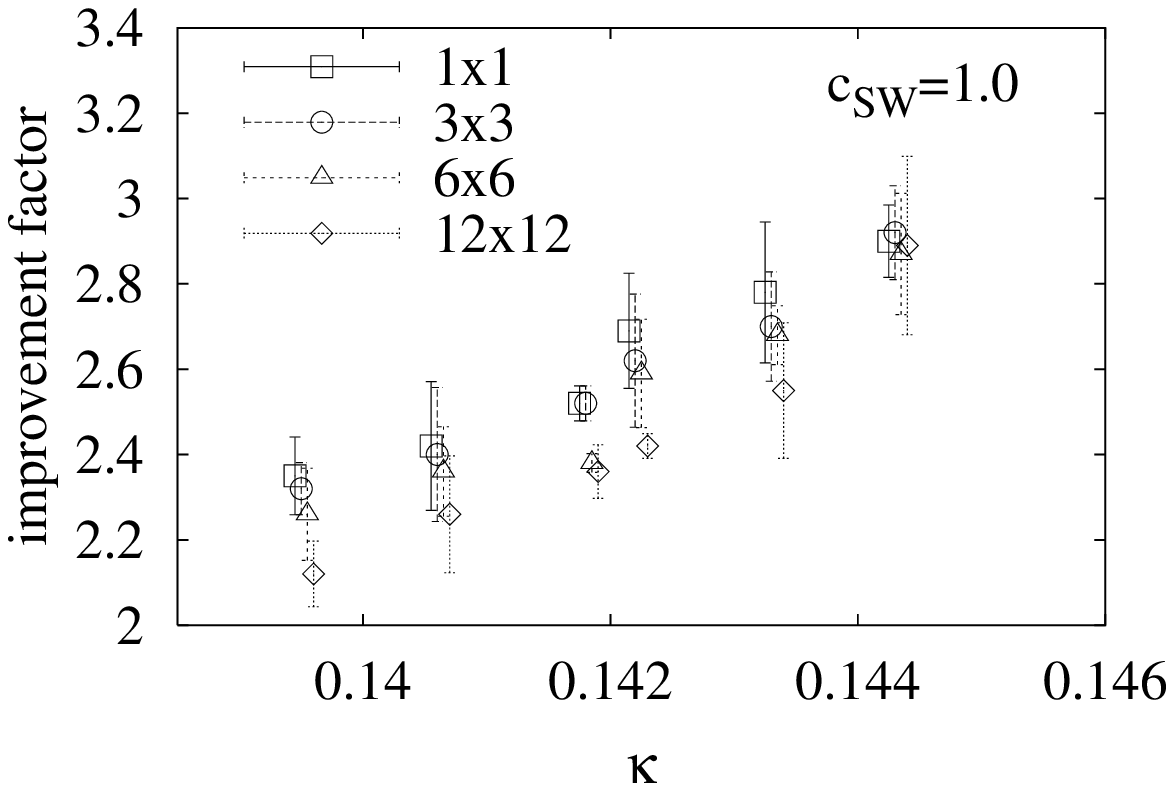}}\\[-.5cm]
{\includegraphics[width=.7\textwidth]{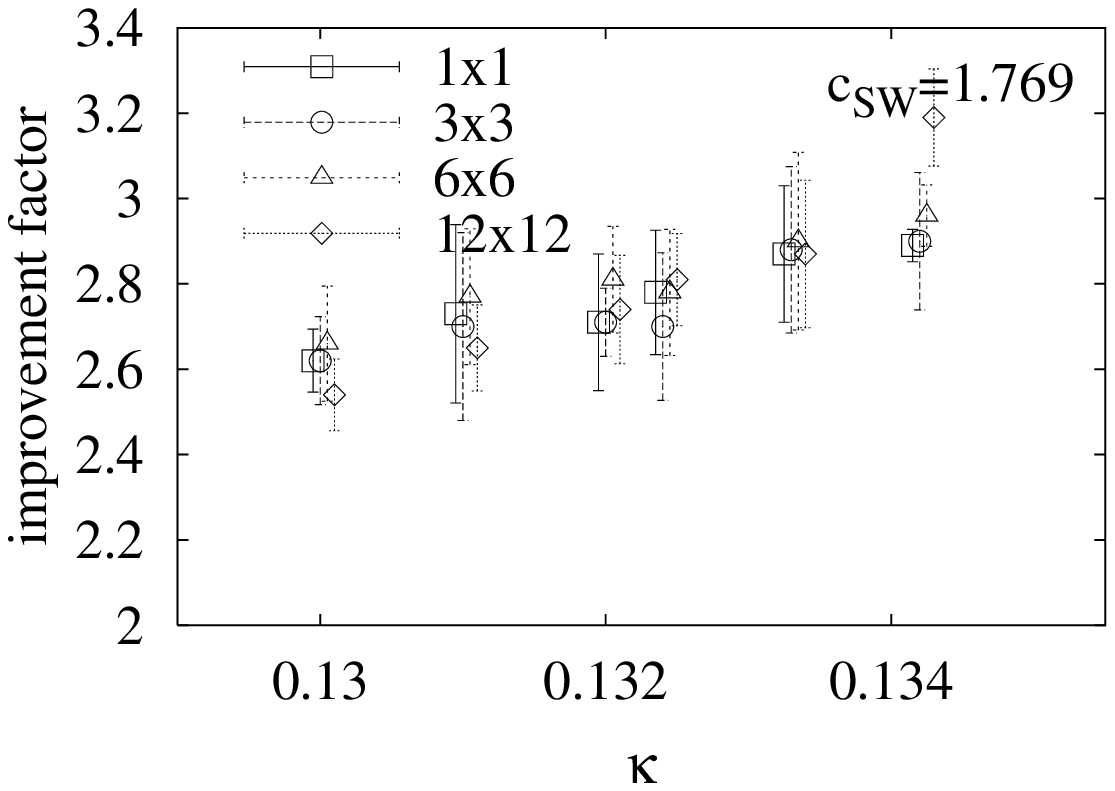}}
\end{center}
\caption{
  $\kappa$ dependence of SSOR peconditioning {\em vs.\ }
  odd-even preconditioning for three values of $c_{SW}$.
\label{OEOYANAGI}}
\end{figure}

Going to the most effective $p=1$ limit, in fact, the \llssor\ 
preconditioner is identical to the SSOR preconditioner which, for
$\omega=1$, is equivalent\footnote{For Wilson parameter $r=1$.}  to an
incomplete LU preconditioning,
 introduced by Oyanagi \cite{Oy85}.  In
\fig{OEOYANAGI}, we present the ensuing improvement factor for the
iteration numbers as a function of $\kappa$ for the three values of
$c_{SW}$, again plotting the gain factor between \llssor\ and odd-even
results ($\omega=1.4$).

As already demonstrated, the gain-factor of SSOR compared to the
odd-even preconditioned system is larger here than for parallel
\llssor.  Close to $\kappa_c$, we can verify a factor of up to 3 for
iteration numbers with increasing tendency going towards $\kappa_c$.
Thus, the improvement factors reported in Refs.~\cite{Oy85} and
\cite{EICKER97} for standard Wilson fermions are confirmed for SWA.
Again, as to the dependence of the improvement factor on $c_{SW}$, one
cannot find a significant variation.

Finally, on the APE100/Quadrics parallel system \cite{QUAD94},
we have implemented and optimized both \llssor\ and odd-even
preconditioners in order to compare real costs.

Following the above results, we measured at $\omega=1.4$,
$\kappa=0.1333$ and $c_{SW}=1.769$.  In the case of \llssor\ we
applied the local diagonal block procedure for $1\times 1$, $3\times
3$, and $6\times 6$ blocks. Additionally, as for the odd-even
preconditioner, we performed a pre-inversion of the local $12\times
12$ blocks.  However, we remark that these blocks require a memory
expense of a factor of 9 compared to the non-blocked version.

\begin{figure}
  \centerline{\includegraphics[width=.7\textwidth]{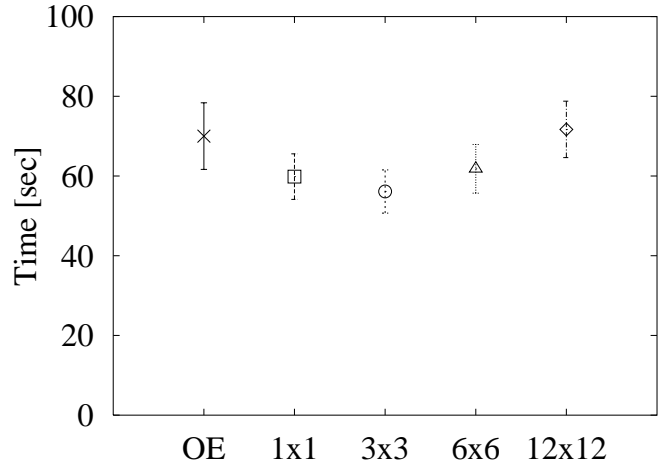}}
\caption{
  Time measurements for the \llssor\ preconditioner on a 32-node
  APE100/Quadrics for SWA. Four different local diagonal blocking methods are
  benchmarked.
\label{TIMESSOR}}
\end{figure}

\begin{figure}
  \centerline{\includegraphics[width=.7\textwidth]{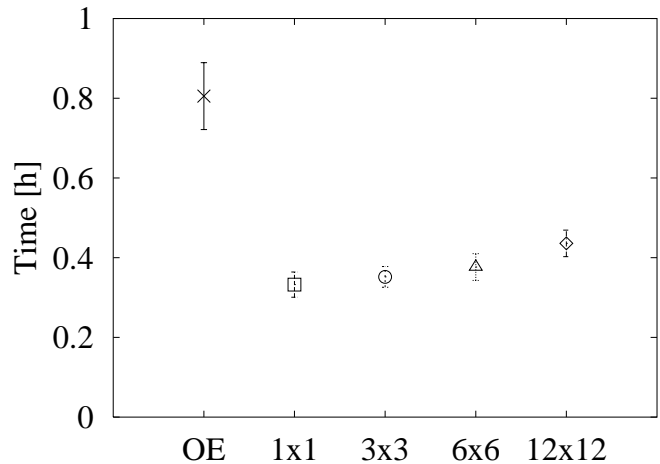}}
\caption{
  Time measurements for the SSOR preconditioner compared to odd-even
  preconditioning on a SUN Ultra workstation.  Again, four different local
  diagonal block size methods are investigated together with the
  odd-even preconditioned system.
\label{TIMESUN}}
\end{figure}

Although one does not achieve an improvement in iteration numbers
between $1\times 1$ and $12\times 12$ blocks, it turns out to be
advantageous to choose a specific block size for a given
implementation machine.  For APE100, the optimal block size is a
$3\times 3$ block.  The results are plotted in \fig{TIMESSOR}.  The
corresponding results as achieved on a SUN Ultra are given in
\fig{TIMESUN}.

On the scalar system, the gain factor in iteration numbers fully pays
off for the diagonal $1\times 1$ algorithm. It is not surprising that
the gain becomes smaller for the larger diagonal blocks since larger
blocks means more compute operations.  The application of the SSOR
principle to solve the system---including the non-trivial
diagonal---completely and thus avoiding its explicit pre-inversion,
turns out to be most effective.  We note that on scalar machines the
gain in iteration numbers fully translates into a gain in compute
time.  On APE100/Quadrics, the improvement gain is deteriorated by
intensive integer operations, a weak point of APE100. 

\subsection{Hypercube Fermions}

The HF action has been rendered gauge invariant `by hand' similar to
the procedures in Ref.~\cite{BIE96,OR97,DEG98}. As such it is,
strictly speaking, not even truncated perfect, however, its coupling
structure is similar to the structure of the latter.

So far we have implemented and tested SSOR preconditioning for HF on a
scalar machine. We have already mentioned that the number of SU(3)
matrices per site to be stored is increased by a factor of 5 compared
to clover fermions.  Limited by the number of hyper-links to store, we
decided to investigate HF on a lattice of size $8^4$. Our
implementation on a SUN Ultra has been written in Fortran90.

\begin{figure}
\begin{center}
\includegraphics[width=.7\textwidth]{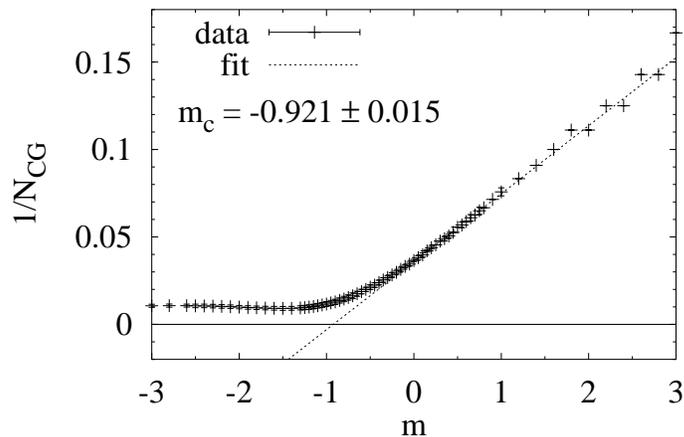}
\end{center}
\caption{
  Determination of $m_c$ by use of CG inversions.
\label{PERFECT:KC}}
\end{figure}

We measured at $\beta=6.0$ in quenched QCD.  First, we tried to assess
the critical mass parameter, in order to determine the critical region
of HF. We used a method introduced in Ref.~\cite{BA92} which makes
use of the dependence of CG iterations on the condition number of the
matrix, cf.\ \fig{PERFECT:KC}.

\begin{figure}
\begin{center}
  \includegraphics[width=.7\textwidth]{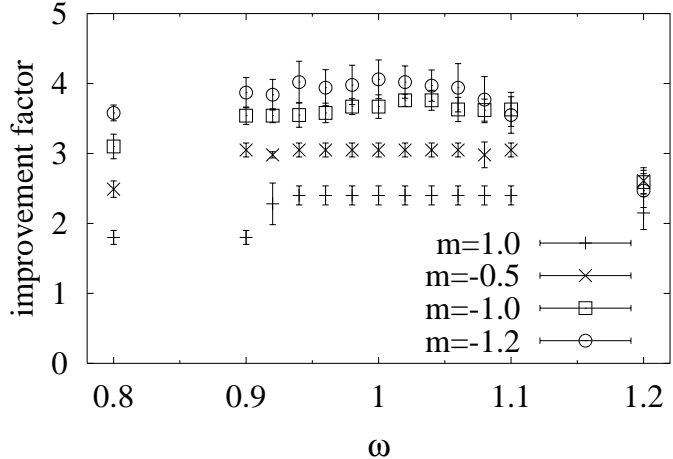}
\end{center}
\caption{
  Dependence of the iteration numbers on the over-relaxation parameter
  $\omega$.
\label{PERFECT:OM}}
\end{figure}

\begin{figure}
\begin{center}
{\includegraphics[width=.7\textwidth]{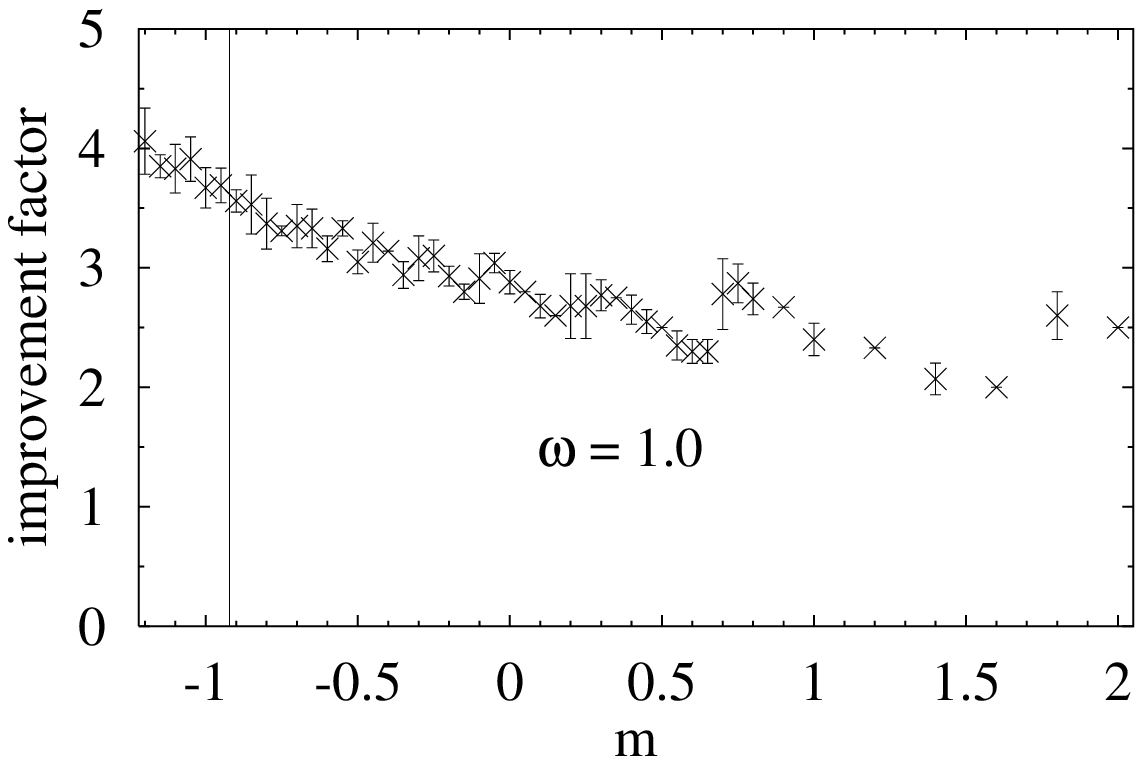}}\\[-.5cm]
{\includegraphics[width=.7\textwidth]{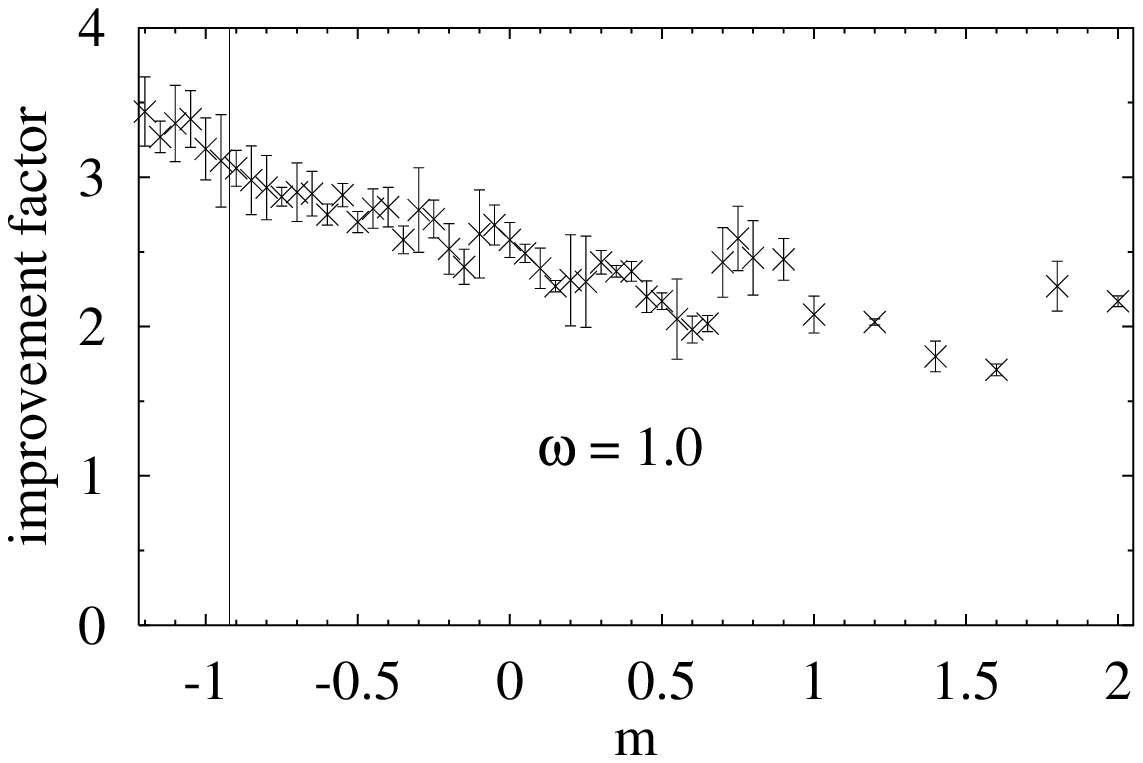}}
\end{center}
\caption{
  Dependence of the solution on the mass parameter $m$.  We show both
  (a) the iteration numbers of the non-preconditioned case {\em vs.\/}
  the \llssor\ preconditioned results and (b) the corresponding gain
  factor in time.
\label{PERFECT:DEP}}
\end{figure}

Close to the critical mass $m_c$ we have assessed the optimal
over-relaxation parameter $\omega$ for the SSOR method, see
\fig{PERFECT:OM}.  Here and in the following we compare the result
with the unpreconditioned BiCGStab solution since simple odd-even
preconditioning cannot be applied in the case of HF.  As a result, we
chose $\omega=1.0$.  \fig{PERFECT:DEP} presents our findings for the
iteration numbers as a function of $m$.

So far, we conclude that SSOR preconditioning of the HF action, gauged
by hand, leads to gain factors up to 4 close to the critical mass
parameter $m_{c}$.  However, we regard the specific form of HF as a
preliminary test case only since a consistent derivation of TPA
includes clover-leaf like terms.

In view of the tremendous compute effort, exceeding that of Wilson
fer\-mions by a factor of more than 10, we regard preconditioning as a
mandatory prerequisite for HF to become competitive with traditional
fermion discretizations.  This conclusion applies equally well to
other variants of HF.

As far as the storage requirements are concerned, again a factor of 10
is found which cannot be avoided. For a given memory limit, this
translates into a factor of 0.56 in linear lattice size or 1.8 in
scale compared to Wilson fermions.  Thus, from a technical point of
view, HF would in principle pay off if their reduced scaling
violations allow coarsening by a factor of 1.8. This is a question for
future investigations where we will include a perfect truncated
gluonic action together with a full-fledged TPA.

\section{Conclusions}

We have constructed {\em locally-lexicographic SSOR preconditioners}
for inversions of linear systems of equations from two improved
fermionic actions, the Sheik\-holeslami-Wohlert-Wilson scheme with
non-constant block-diagonal and re\-normalization group improved
hypercube fermions, with interaction gauged {\em ad hoc}.

For SWA we find the block \llssor-scheme to be more effective by
factors up to 2.5 compared to odd-even preconditioned solvers.  For HF
we have demonstrated that SSOR preconditioning accelerates the
iterative solver by a factor of 3 to 4 compared to the
non-preconditioned system. We believe that the improvement for HF will
translate also into other TPA with interaction derived from
renormalization group transformations.


\begin{thebibliography}{99}

\frenchspacing
 
\bibitem{GUP95} T. Bhattacharya, R. Gupta, G. Kilcup, and S. Sharpe:
  Phys. Rev. D53 (1996) 6486.
  
\bibitem{LEP95} G.P. Lepage: Nucl. Phys. B (Proc. Suppl.) 47 (1996) 3.
  
\bibitem{NIE96} F. Niedermayer: Nucl. Phys. B (Proc. Suppl.) 53 (1997) 56.
  
\bibitem{GUP96} R. Gupta and T. Bhattacharya: Phys. Rev. D55 (1997)
  7203.
  
\bibitem{JAN95} K. Jansen, C. Liu, M. L\"{u}scher, H. Simma, S. Sint,
R. Sommer, P. Weisz, and U. Wolff: Phys. Lett. B372 (1996) 275.

\bibitem{JAN98} K. Jansen and R. Sommer: {\it O(a) improvement of
    lattice QCD with two flavors of Wilson quarks}, preprint 
  hep-lat/9803017.
  
\bibitem{FRO94} A. Frommer, V. Hannemann, B. N\"ockel, Th. Lippert, and
  K. Schilling: Int. J.  of Mod. Phys. C Vol.~5 No.~6 (1994) 1073.
  
\bibitem{FIS96} S. Fischer, A. Frommer, U. Gl\"{a}ssner, Th. Lippert,
  G.  Ritzenh\"{o}fer and K. Schilling: Comp. Phys. Comm. 98 (1996)
  20.

\bibitem{SHE85} B. Sheikholeslami and R. Wohlert: Nucl. Phys. B259
  (1985) 572.

\bibitem{SYM83} K. Symanzik: Nucl. Phys. B212 (1983) 1.
  
\bibitem{LUE962} M. L\"{u}scher, S. Sint, R. Sommer, and P. Weisz:
  Nucl. Phys. B478 (1996) 365.
  
\bibitem{Lue97} M. L\"uscher, S. Sint, R. Sommer, P.  Weisz, and U.
  Wolff: Nucl. Phys. B 491 (1997) 323.
  
\bibitem{WiKo} K. Wilson and J. Kogut: Phys. Rep. C12 (1974) 75.\\
  K.  Wilson: Rev. Mod. Phys. 47 (1975) 773.

\bibitem{HAS94} P. Hasenfratz and F. Niedermayer: Nucl. Phys. 
B414 (1994) 785.

\bibitem{BIE96} W. Bietenholz, R. Brower, S. Chandrasekharan, and U.-J.
  Wiese: Nucl. Phys. B (Proc. Suppl.) 53 (1997) 921.
  
\bibitem{OR97} K. Orginos, W. Bietenholz, R. Brower, S.
  Chandrasekharan, and U.-J.  Wiese: Nucl. Phys. B (Proc. Suppl.) 63
  (1998) 904.
  
\bibitem{DEG98} MILC Collaboration: T. DeGrand, {\it Tests of
    Hypercubic Fermion Actions}, preprint hep-lat/9802012.
  
\bibitem{BA92} I. Barbour, E. Laermann, Th. Lippert, and K. Schilling:
  Phys. Rev. {\bf D 46} (1992) 3618.

\bibitem{BW96} W. Bietenholz and U.-J. Wiese: Nucl. Phys.
B464 (1996) 319.

\bibitem{BIE98} W. Bietenholz and U.-J. Wiese: Phys. Lett. B426 (1998)
  114.
  
\bibitem{Eis} S. Eisenstat: SIAM J. Sci. Stat. Comput. 2 (1981) 1.

\bibitem{JAN96} K. Jansen and C. Liu: Comp. Phys. Comm. 99 (1997)
  221.
  
\bibitem{QUAD94} C. Battista et al.: Int. J. of High Speed Computing 5 (1993) 637.

\bibitem{Oy85} Y. Oyanagi: Comp.  Phys. Comm. 42 (1986) 333.
  
\bibitem{EICKER97} N. Eicker, W. Bietenholz, A. Frommer, H. Hoeber,
  Th. Lippert, and K. Schilling: Nucl. Phys. B (Proc. Suppl.) 63
  (1998) 955.
\end{thebibliography}
\end{document}